\documentclass[sigconf, balance=false, screen]{acmart}
\usepackage{style-Files/acmart/popets}

\setcopyright{popets}
\copyrightyear{YYYY}
\acmYear{YYYY}
\acmVolume{YYYY}
\acmNumber{X}
\acmDOI{XXXXXXX.XXXXXXX}
\acmISBN{}
\acmConference{Proceedings on Privacy Enhancing Technologies}
\usepackage{dsfont}
\usepackage{tikz}
\usepackage{multirow}
\usepackage{comment}
\usepackage{url}
\usepackage{mathtools}
\usepackage{xcolor}
\usepackage{hyperref}
\usepackage{colortbl}

\usepackage{hyperref}
\def\authnote{1}
\usepackage{booktabs}
\usepackage{xcolor}
\usepackage{subcaption}
\usepackage{footmisc}
\usepackage{newtxmath}
\usepackage{nicefrac}
\usepackage{mathtools}% Loads amsmath
\usepackage{bm}
\usepackage{algpseudocode}

\usepackage[linesnumbered,boxed]{algorithm2e}
\usepackage{fullwidth}
\usepackage{etoolbox,xspace}
\usepackage{xurl}
\usepackage{physics}

\newcommand{\secref}[1]{\mbox{Section~\ref{#1}}}
\newcommand{\eqnref}[1]{\mbox{Equation~(\ref{#1})}}

\newcommand{\funcfont}[1]{\textsf{#1}}
\usepackage{adjustbox}

\usepackage{bbding}
\usepackage{pifont}
\usepackage{wasysym}

\usepackage[detect-all]{siunitx}
\usepackage[utf8]{inputenc}
\usepackage{pgfplots}
\DeclareUnicodeCharacter{2212}{−}
\usepgfplotslibrary{groupplots,dateplot}
\usetikzlibrary{patterns,shapes.arrows}
\pgfplotsset{compat=newest}
\usepackage{tablefootnote}

\usepackage{booktabs}
\usepackage{bigfoot}
\DeclareNewFootnote{AAffil}[arabic]
\DeclareNewFootnote{ANote}[fnsymbol]

\usepackage{etoolbox}
\makeatletter
\patchcmd\maketitle{\def\@makefnmark{\rlap{\@textsuperscript{\normalfont\@thefnmark}}}}{}{}{}
\makeatother

\makeatletter
\def\thanksAAffil#1{% <--- These %'s are necessary for spacing
  \footnotemarkAAffil\protected@xdef\@thanks{\@thanks%
        \protect\footnotetextAAffil[\the \c@footnoteAAffil]{#1}}%
}
\def\thanksANote#1{%
  \footnotemarkANote%
  \protected@xdef\@thanks{\@thanks%
        \protect\footnotetextANote[\the \c@footnoteANote]{#1}}%
}
\makeatother

\newcommand{\sparagraph}[1]{\medskip\noindent\textbf{#1}\ }

\newcommand{\bnm}{\begin{newmath}}
\newcommand{\enm}{\end{newmath}}

\newcommand{\bea}{\begin{neweqnarrays}}%
\newcommand{\eea}{\end{neweqnarrays}}%

\newcommand{\bne}{\begin{newequation}}
\newcommand{\ene}{\end{newequation}}

\newcommand{\bal}{\begin{newalign}}
\newcommand{\eal}{\end{newalign}}

\newenvironment{newalign}{\begin{align*}%
\setlength{\abovedisplayskip}{4pt}%
\setlength{\belowdisplayskip}{4pt}%
\setlength{\abovedisplayshortskip}{6pt}%
\setlength{\belowdisplayshortskip}{6pt} }{\end{align*}}

\newenvironment{newmath}{\begin{displaymath}%
\setlength{\abovedisplayskip}{4pt}%
\setlength{\belowdisplayskip}{4pt}%
\setlength{\abovedisplayshortskip}{6pt}%
\setlength{\belowdisplayshortskip}{6pt} }{\end{displaymath}}

\newenvironment{neweqnarrays}{\begin{eqnarray*}%
\setlength{\abovedisplayskip}{4pt}%
\setlength{\belowdisplayskip}{4pt}%
\setlength{\abovedisplayshortskip}{4pt}%
\setlength{\belowdisplayshortskip}{4pt}%
\setlength{\jot}{0.0in} }{\end{eqnarray*}}

\newenvironment{newequation}{\begin{equation}%
\setlength{\abovedisplayskip}{4pt}%
\setlength{\belowdisplayskip}{4pt}%
\setlength{\abovedisplayshortskip}{6pt}%
\setlength{\belowdisplayshortskip}{6pt} }{\end{equation}}

\newcounter{ctr}

\newenvironment{newitemize}{%
\begin{list}{\mbox{}\hspace{5pt}$\bullet$\hfill}{\labelwidth=15pt%
\labelsep=4pt \leftmargin=12pt \topsep=3pt%
\setlength{\listparindent}{\saveparindent}%
\setlength{\parsep}{\saveparskip}%
\setlength{\itemsep}{3pt} }}{\end{list}}

\newenvironment{newenum}{%
\begin{list}{{\rm (\arabic{ctr})}\hfill}{\usecounter{ctr} \labelwidth=17pt%
\labelsep=3pt \leftmargin=21pt \topsep=3pt%
\setlength{\listparindent}{\saveparindent}%
\setlength{\parsep}{\saveparskip}%
\setlength{\itemsep}{2pt} }}{\end{list}}

\newlength{\saveparindent}
\newlength{\saveparskip}
\newcommand{\E}{{\rm I\kern-.3em E}}

\newcommand{\given}{\ensuremath{\,\big|\,}}

\newcommand{\appref}[1]{\mbox{Appendix~\ref{#1}}}

\newcommand{\figref}[1]{\mbox{Figure~\ref{#1}}}
\renewcommand{\eqref}[1]{\mbox{Equation~(\ref{#1})}}
\newcommand{\tabref}[1]{\mbox{Table~\ref{#1}}}

\newcommand{\get}{{\:{\leftarrow}\:}}
\newcommand{\gett}[1]{{\:{\leftarrow{\hspace*{-3pt}\raisebox{.75pt}{$\scriptscriptstyle #1$}}}\:}}
\newcommand{\getsr}{\gett{\$}}

\newcommand{\gamesfontsize}{\footnotesize}
\newcommand{\tabfontsize}{\footnotesize}

\providecommand{\setstretch}[1]{}
\newcommand{\mpage}[2]{\begin{minipage}[t]{#1\textwidth}\setstretch{1.03}\gamesfontsize\vspace{0pt}  #2 \end{minipage}}% \begin{minipage}[t]{#1\textwidth}\setstretch{1.03}\gamesfontsize  #2 \end{minipage}
\newcommand{\framedminipage}[2]{\framebox{\mpage{#1}{#2}}}
\newcommand{\fpage}[2]{\framebox{\mpage{#1}{#2}}}

\newcommand{\hpagess}[4]{
        \begin{tabular}[t]{c@{\hspace*{.5em}}c}
        \begin{minipage}[t]{#1\textwidth}\gamesfontsize #3 \end{minipage}
        &
        \begin{minipage}[t]{#2\textwidth}\gamesfontsize #4 \end{minipage}
        \end{tabular}
    }

\newcommand{\hfpagess}[4]{
  \begin{tabular}[t]{c@{\hspace*{.2em}}c}
    \framedminipage{#1}{#3}
    & \framedminipage{#2}{#4}
  \end{tabular}
}

\newcommand{\Z}{\mathbb{Z}}

\def \part {part}

\def \adv {{\mathcal A}}

\DeclareMathOperator*{\argmin}{argmin}

\renewcommand{\paragraph}[1]{\vspace*{6pt}\noindent\textbf{#1}\;}

\newcounter{mytable}
\def\mytable{\begin{centering}\refstepcounter{mytable}}
\def\endmytable{\end{centering}}

\newcounter{myfig}
\def\myfig{\begin{centering}\refstepcounter{myfig}}
\def\endmyfig{\end{centering}}

\def \blackslug{\hbox{\hskip 1pt \vrule width 4pt height 8pt
    depth 1.5pt \hskip 1pt}}
\def \qed{\quad\blackslug\lower 8.5pt\null\par}
\DeclareMathOperator{\poly}{\mbox{$f$}}

\newcommand\ignore[1]{}

\newcounter{mynote}[section]
\newcommand{\notecolor}{blue}
\newcommand{\thenote}{\thesection.\arabic{mynote}}
\newcommand{\tnote}[1]{\ifnum\authnote=1\refstepcounter{mynote}{\bf \textcolor{\notecolor}{$\ll$TomR~\thenote: {\sf #1}$\gg$}}\fi}

\newcommand{\better}[1]{\ifnum\authnote=1{\textcolor{violet}{[BetterWord: #1]}}\fi}
\newcommand{\todo}[1]{\ifnum\authnote=1{\textcolor{blue}{[TODO: #1]}}\fi}

\newcommand{\dnote}[1]{\ifnum\authnote=1\refstepcounter{mynote}{\bf \textcolor{red}{$\ll$DM~\thenote: {\sf #1}$\gg$}}\fi}
\newcommand{\perinote}[1]{\ifnum\authnote=1\refstepcounter{mynote}{\bf \textcolor{red}{$\ll$Peri~\thenote: {\sf #1}$\gg$}}\fi}
\newcommand{\mnote}[1]{\ifnum\authnote=1\refstepcounter{mynote}{\bf \textcolor{\notecolor}{$\ll$Mazharul~\thenote: {\sf #1}$\gg$}}\fi}
\newcommand{\rcnote}[1]{\ifnum\authnote=1{\bf \textcolor{magenta}{$\ll$RC: {\sf #1}$\gg$}}\fi}
\newcommand{\mytab}{\hspace*{.4cm}}

\DeclareMathSymbol{\mlq}{\mathord}{operators}{``}
\DeclareMathSymbol{\mrq}{\mathord}{operators}{`'}

\newcommand{\rhf}[2]{R_{f, \gamma}}

\def\ret{\ensuremath{\mathbf{return}}\xspace}
\newcommand{\olrk}[1]{\ifx\nursymbol#1\else\!\!\mskip4.5mu plus 0.5mu\left (\mskip0.5mu plus0.5mu #1\mskip1.5mu plus0.5mu \right)\fi}

\newcommand{\floor}[1]{\left \lfloor #1 \right \rfloor }

\newcommand{\indicator}{I}

\def\normal{\mathcal{N}}

\newcommand{\abythree}{ABY3\xspace}

\newcommand{\sigmoid}{\ensuremath\mathsf{sigmoid}}

\newcommand{\relu}{\ensuremath\funcfont{ReLU}}
\newcommand{\silu}{\ensuremath\funcfont{SiLU}}
\newcommand{\mish}{\ensuremath\funcfont{Mish}}
\newcommand{\gelu}{\ensuremath\funcfont{GeLU}}
\newcommand{\tanhh}{\ensuremath\funcfont{TanH}}

\renewcommand{\deg}{\ensuremath \textsf{Deg}}
\newcommand{\maxdegree}{\ensuremath \mathsf{k}}
\newcommand{\density}{\ensuremath{\textsf{P}}}
\newcommand{\npieces}{\ensuremath \mathsf{m}}

\newcommand{\ethreshold}{\ensuremath \delta}
\newcommand{\ecrude}{\ensuremath \delta'}
\newcommand{\sysname}{\textsf{Compact}\xspace}
\newcommand{\nfgen}{\textsf{NFGen}\xspace}

\DeclareMathOperator{\Error}{\ensuremath \mathcal{E}}
\newcommand{\MaxApproxError}{\ensuremath \Error_{\mathsf{Max}}}
\newcommand{\MeanApproxError}{\ensuremath \Error_{\mathsf{Mean}}}

\newcommand{\Time}{\ensuremath \funcfont{Time}}
\newcommand{\AccuracyLoss}{\ensuremath \funcfont{AccLoss}}
\newcommand{\randomNumber}{\ensuremath r}
\newcommand{\negligibleaccuracy}{\ensuremath \nu}

\newcommand{\comments}[1]{{\textit{\textcolor{gray}{ // \texttt{#1} }}}}
\newcommand{\fx}{\ensuremath F_{\mathsf{act}}}
\newcommand{\fmpc}{\ensuremath \widehat{F}_{\mathsf{act}}}
\newcommand{\fcur}{\ensuremath \fmpc^{\funcfont{cur}}}
\newcommand{\tcur}{\ensuremath \theta_{\funcfont{cur}}}

\newcommand{\fcrude}{\ensuremath \widehat{F}_\textrm{\sf act}^{\textrm{\sf crd}}}
\newcommand{\newfcrude}[1]{\ensuremath \widehat{F}_\mathsf{#1}^{\textsf{crd}}}

\newcommand{\stepsize}{\ensuremath \Delta}
\newcommand{\step}{ \ensuremath i}

\newcommand{\interpolate}{\ensuremath \funcfont{InterPolate}}

\newcommand{\server}{\ensuremath \mathcal{S}}
\newcommand{\modelowner}{\ensuremath \server_\textsf{owner}}
\newcommand{\client}{\ensuremath \mathcal{C}}

\newcommand{\ApproximationAlgo}{\ensuremath \funcfont{FindBestPiecePoly}\xspace}
\newcommand{\candidatelistgenerationalgo}{\ensuremath \funcfont{GenAccuracteApprox}\xspace}
\newcommand{\generateNeighbour}{\ensuremath \funcfont{GenerateNeighbour}}

\newcommand{\ring}{\ensuremath \mathcal{R}}

\newcommand{\addmathoperation}{\ensuremath \funcfont{ADD}}
\newcommand{\comparemathoperation}{\ensuremath \funcfont{COMP}}
\newcommand{\mulmathoperation}{\ensuremath \funcfont{MUL}}

\newcommand{\taskA}{ \textsf{DigitRecognition}\xspace}
\newcommand{\taskB}{ \textsf{CIFAR10Classification}\xspace}
\newcommand{\taskC}{  \textsf{ImageNet1KClassification}\xspace}
\newcommand{\taskD}{ \textsf{SpoofFaceDetection}\xspace}

\colorlet{electriclime}{gray!20}

\newcommand{\clientinput}{\ensuremath \textbf{x}}
\newcommand{\weights}{\ensuremath \textbf{w}}
\newcommand{\modelinput}{\ensuremath \textbf{W}}

\newcommand{\solution}{\ensuremath \theta}

\newcommand{\plainTextAcc}{\ensuremath \eta}
\newcommand{\secureInfAcc}{\ensuremath \plainTextAcc'}

\let\oldnl\nl% Store \nl in \oldnl
\newcommand{\nonl}{\renewcommand{\nl}{\let\nl\oldnl}}% Remove line number for one line

\newcommand*\emptycirc[1][1ex]{\tikz\draw (0,0) circle (#1);} 
\newcommand*\halfcirc[1][1ex]{%
	\begin{tikzpicture}
	\draw[fill] (0,0)-- (90:#1) arc (90:270:#1) -- cycle ;
	\draw (0,0) circle (#1);
	\end{tikzpicture}}
 
\newcommand*\fullcirc[1][1ex]{\tikz\fill (0,0) circle (#1);}
\newcommand{\changed}[1]{{\ifnum\authnote=1 {\color{black} #1} \else #1 \fi}}
\newcommand*\circled[1]{\tikz[baseline=(char.base)]{
            \node[shape=circle,fill,inner sep=1pt] (char) {\textcolor{white}{#1}};}}
\usepackage{enumitem}

\newcommand{\resolution}{\ensuremath \ell}
\newcommand{\decimal}{\ensuremath d}

\def\thetitle{$\sysname$: 
Approximating Complex Activation Functions for Secure Computation}

\title{\thetitle}

\begin{document}
%-------------------------------------------------------------------------------
% \author{
%   \IEEEauthorblockN{%
%     Mazharul Islam\IEEEauthorrefmark{1}\textsuperscript{\textsection},
%     Sunpreet S. Arora\IEEEauthorrefmark{2},
%     Rahul Chatterjee\IEEEauthorrefmark{1}, 
%     Peter Rindal\IEEEauthorrefmark{2},
%     Maliheh Shirvanian\IEEEauthorrefmark{3}\textsuperscript{\textsection}%
%   }%
%   \IEEEauthorblockA{\IEEEauthorrefmark{1} University of Wisconsin-Madison}%
%   \IEEEauthorblockA{\IEEEauthorrefmark{2} Visa Research}%
%   \IEEEauthorblockA{\IEEEauthorrefmark{3} Netflix}%
%   }
% \author{
%   Anonymous Author(s)
% }
% \author{Mazharul Islam$^\dag$, $^\ddag$, $^\dag$, Peter Rindal$^\ddag$, Maliheh Shirvanian$^\S$ \\ 
% University of Wisconsin-Madison$^\dag$, Visa Research$^\ddag$, Netflix$ ^\S$
% someemail@somedomain.com
% }
% \and
% \IEEEauthorblockN{Homer Simpson}
% \IEEEauthorblockA{Twentieth Century Fox\\
% homer@thesimpsons.com}
% \and
% \IEEEauthorblockN{James Kirk\\ and Montgomery Scott}
% \IEEEauthorblockA{Starfleet Academy\\
% someemail@somedomain.com}}

\author{Mazharul Islam$^{\ast}$, Sunpreet S. Arora$^\dag$, Rahul Chatterjee$^{\ast}$, Peter Rindal$^\dag$, Maliheh Shirvanian$^{\ddagger}$\vspace{1mm}}
\affiliation{
    \institution{$^{\ast}$ University of Wisconsin---Madison,\hspace{1em} $^\dag$ Visa Research, $^\ddagger$ Netflix}
}
% \authornote{Work partially done when first, and last authors were both at Visa Research.}
% \email{mislam9@wisc.edu}
% \authornote{Work done while  author was an intern at Visa Research}

% \author{Sunpreet S. Arora}
% \affiliation{
%     \institution{ Visa Research}
% }
% \email{sunarora@visa.com}

% \author{Rahul Chatterjee}
% \affiliation{
%     \institution{University of Wisconsin - Madison}
% }
% \email{rchatterjee4@wisc.edu}

% \author{Peter Rindal}
% \affiliation{
%     \institution{Visa Research}
% }
% \email{perindal@visa.com}

% \author{Maliheh Shirvanian}
% \affiliation{
%     \institution{Netflix}
% }
% \email{maliheh21@gmail.com}
% \authornote{Work done while  author was at Visa Research}

\renewcommand{\shortauthors}{Islam et al.}

% <concept_id>10002978.10002991.10002995</concept_id>
% <concept_desc>Security and privacy~Privacy-preserving protocols</concept_desc>

% \ccsdesc[500]{Security and privacy~Privacy-preserving protocols}
% \keywords{Deep neural networks, Complex activation functions, Secure evaluation}

% \IEEEoverridecommandlockouts
% \makeatletter\def\@IEEEpubidpullup{6.5\baselineskip}\makeatother
% \IEEEpubid{\parbox{\columnwidth}{
%     Network and Distributed System Security (NDSS) Symposium 2024\\
%     26 February - 1 March 2024, San Diego, CA, USA\\
%     ISBN 1-891562-93-2\\
%     https://dx.doi.org/10.14722/ndss.2024.23xxx\\
%     www.ndss-symposium.org
% }
% \hspace{\columnsep}\makebox[\columnwidth]{}}
\makeatletter
\def\@ACM@checkaffil{% Only warnings
    \if@ACM@instpresent\else
    \ClassWarningNoLine{\@classname}{No institution present for an affiliation}%
    \fi
    \if@ACM@citypresent\else
    \ClassWarningNoLine{\@classname}{No city present for an affiliation}%
    \fi
    \if@ACM@countrypresent\else
        \ClassWarningNoLine{\@classname}{No country present for an affiliation}%
    \fi
}
\makeatother

\makeatletter
\renewcommand{\boxed}[1]{\text{\fboxsep=.2em\fbox{\m@th$\displaystyle #1 $}}}
\makeatother


\begin{abstract}
% Secure multi-party computation (MPC) techniques can be used 
% to provide data privacy when users query deep neural network (DNN) models hosted on a public cloud.
% DNN models consists of \fixme{thousands of} non-linear activation functions (AFs), and computing them efficiently using MPC techniques is challenging.
% This is especially true for  complex and highly non-linear AFs present in DNN models used for cutting-edge applications. 
% % often use
% % Computing non-linear activation functions (AFs) of DNN models using MPC techniques is a major challenge, and 
% Existing work to solve this challenge mostly focus on simple AFs such as $\relu$.  
% Designing efficient MPC techniques for  complex AFs is an open problem.

Secure multi-party computation (MPC) techniques can be used to provide data privacy when users query deep neural network (DNN) models hosted on a public cloud. State-of-the-art MPC techniques can be directly leveraged for DNN models that use simple activation functions such as ReLU. 
However, these techniques are ineffective and/or inefficient for the complex and highly non-linear activation functions used in cutting-edge DNN models. 
%However, DNN model architectures designed for cutting-edge applications often use complex and highly non-linear AFs. Designing efficient MPC techniques for such complex AFs is an open problem.

We present \sysname, which produces piece-wise polynomial approximations of complex activation functions 
that can be used
%to enable their efficient use 
with state-of-the-art MPC techniques. 
% features of our approach
\sysname neither requires nor imposes any restriction on model training and 
achieves%results in 
near-identical model accuracy. 
% what techniques we use to achieve these features
%To achieve this, 
We design \sysname with input density awareness and 
use an application specific simulated annealing type optimization to generate computationally efficient approximations of complex activation functions.
% experimental results.
We extensively evaluate \sysname on four different machine-learning tasks 
with DNN architectures that use popular complex activation functions $\silu$, $\gelu$, and $\mish$. Our experimental results show that $\sysname$ incurs 
negligible accuracy loss 
while being 2$\times$---5$\times$ 
faster %computationally more efficient  
than state-of-the-art approaches for DNN models with large number of hidden layers. 
% Larger implications
Our work 
accelerates easy adoption of MPC techniques to provide user data privacy even when
the queried DNN models consist of a number of hidden layers and 
%trained over 
complex activation functions.
\end{abstract}

\maketitle
% \pagestyle{plain}
% \begingroup\renewcommand\thefootnote{\textsection}
% \footnotetext{Work done while at Visa Research}
% \endgroup
% \thispagestyle{empty}

% \vspace{-0.5cm}
\section{Introduction} \label{sec:intro}
% \fixme{Try to find an anecdote from secure inference from ChatGPT papers. optional}
Deep neural networks (DNNs) based inference services   
are being increasingly adopted in various emerging applications, such as early disease discovery from personal health records~\cite{tomavsev2019clinically}, personalized product recommendations~\cite{recommendations-ai}, media translations~\cite{media-translation}, image recognition~\cite{camera-face}, and even biometric authentication~\cite{irs-to-drop-facial-scan}.
Trained DNN models are typically hosted on a cloud server for applications or users 
to query for inference tasks. These services however can pose serious privacy concerns.
For instance, users are required to share their facial images with an online service 
hosting a face recognition DNN model.
Indeed due to such privacy concerns, the Internal Revenue Service (IRS) removed the identity verification service based on facial recognition~\cite{irs-to-drop-facial-scan}.
% due to the concern of the privacy of users' facial images

DNN models used for inference cannot be transferred to the client devices because they can be proprietary and trained on private training data such as users' medical records~\cite{fredrikson2015model,10.1145/3274694.3274740}.  Clients also would like to avoid sharing their private data with the model hosting server. %  for inference purposes.  
This problem is generally referred to as \emph{secure inference}, where a client can obtain the inference results on their private input without sharing it with the server, nor learning anything about the DNN model parameters.  
% Secure inference can resolve privacy risks associated with DNN-based inference services. 
% Recent progress in 
The secure 
multi-party computation (MPC) is a promising approach
% has resulted in several investigations 
to solve the secure inference problem~\cite{xu2021privacy}. % cite a survey paper 
However, a key challenge is 
computing %handling 
the non-linear activation functions (AFs) efficiently. 
%as there is no efficient MPC technique to compute nonlinear functions. 
Indeed, studies have shown that AFs are the bottleneck --- compared to  
linear layers --- for performing secure inference~\cite{garimella2021sisyphus,  hussain2021coinn, rathee2020cryptflow2, mishra2020delphi, ghodsi2020cryptonas}.

% Thus, a suitable secure inference protocol, needs to handle the AFs in a way such that both accuracy loss and performance overhead are negligible.
Prior works~\cite{rathee2020cryptflow2, chandran2021simc, lehmkuhl2021muse, gilad2016cryptonets, riazi2019xonn, dalskov2020secure} % addressing this problem, 
have provided several solutions for secure inference for DNN models with $\relu$
AF --- a relatively simple non-linear AFs.
However, lately, ML researchers are favoring more  
%and DNN models are being % designed for cutting-edge applications often 
complex and highly non-linear AFs
such as  $\silu, \gelu, \mish$ for new ML applications. State-of-the-art secure inference protocols are either unsuitable or inefficient for  DNN models trained over these complex AFs.

%and further retrain/fine-tune the trained DNN models to make it compatible with existing secure inference protocols specific to $\relu$. However, retraining is expensive, and it is not clear switching to $\relu$ can still retain the robustness, noise-resistant %, and better-performing  properties which complex AFs infuse DNN models with.

\changed{
Complex AFs can be approximated using 
%A standard approach to handle AFs for secure inference is use % a number of 
piece-wise polynomials for efficient computation in MPC frameworks, as shown in recent works~\cite{liu2017oblivious,li2022mpcformer,sirnn-lib}. 
However, a key limitation of these approaches is that they incur high accuracy loss compared to plaintext (``not secure'') inference.   
Fan et al. recently proposed \nfgen~\cite{NFGenFanCCS22}, that can generate MPC-friendly polynomial approximation of a variety of non-linear functions used in scientific domains without introducing significant accuracy loss.  
Although this approach is generic and can be used to approximate complex AFs, doing so with \nfgen incurs significant performance overhead (as we show in~\secref{sec:performance}). 

This is because generating approximations of complex AFs that have both negligible accuracy loss, and performance overhead requires carefully searching for optimal parameters used in the approximation process.
Conservatively setting these parameters, 
as done in %similar to 
\nfgen, to generate an approximation that does not introduce significant accuracy loss will in turn increase the computational overhead. 
One may settle for parameters that yield an imprecise approximation of AFs to improve the speed of secure inference using \nfgen, but it will end up degrading the accuracy of the DNN model. 
Swapping out the complex AF with $\relu$ increases accuracy loss, and might require retraining or fine-tuning, which are computationally expensive for large DNN models. 
}

We devise a new approach to approximate complex AFs that does not degrade inference accuracy even for DNN models with many hidden layers.  We do so without requiring any retraining of the model or change in the model architecture.  

There are two main challenges to achieve an improved approximation of AFs. 
First,  due to input normalization, 
most of the inputs to AFs are 
around zero, 
%close to zero due to input normalization and 
where complex AFs are highly non-linear. 
%near zero. 
\changed{Chabanne et al.~\cite{chabanne2017privacy} observed that for a nine-layered DNN model normalization pushes $99.73\%$ of the input values to the $\relu$ AF between $[-3, 3]$.}  Second, the approximation approach needs to balance the trade-offs between performance overhead and inference accuracy loss carefully.

To handle these two challenges, we incorporate the observation that in state-of-the-art DNN models, inputs to complex AFs are normalized, into our
approximation generation process. 
Such normalization gives a way to estimate the input probability density to the complex AF as  
the majority of the normalized inputs to the complex AF would fall into specific places near the region close to zero with high probability --- while a small portion will fall into places in regions away from zero with low probability.
We hypothesize that taking this observation into account will help mitigate the cumulative impact of errors introduced by MPC-friendly approximations from one layer to subsequent layers of a deeper or wider DNN model.
\changed{ While prior work~\cite{chabanne2017privacy} has used this observation for generating fully homomorphic encryption (FHE)-friendly approximations of $\relu$ AF, their proposed approach of using 
a single polynomial generated via ``least square fit''  would not work well for MPC-friendly approximations of complex AFs when DNN models have a high number of hidden layers~\cite{miranda1996lecture}.} 
% Using these two techniques, we design a scheme that finds the Given $\npieces, \maxdegree, \ring$,

We use the Chebyshev sequence-based interpolation~\cite{chebinterpolation} for piece-wise polynomial approximation, which is shown to provide better approximation for non-linear functions involving operations, including $e^{-x}, \ln(x), \tanh(x), 1/x$, etc., as it is the case with complex AFs~\cite[Table 5.2]{miranda1996lecture}.~\footnote{Although \nfgen also used Chebyshev interpolation, they did not consider \\ input normalization to improve accuracy.} 

% by harmonizing the intricate architecture of the state-of-the-art DNN models into the approximation process 
% while maintaining compatibility with existing general-purpose MPC libraries. 
% first, similar to prior work~\cite{NFGenFanCCS22}, we use the Chebyshev sequence-based interpolation~\cite{chebinterpolation} for piece-wise polynomial approximation. 
% which is more suitable for generating MPC-friendly approximation of complex non-linear functions. 
% Other interpolation techniques --- relied on by prior works such as cubic spline-based~\cite{liu2017oblivious}, least square fit piece-wise interpolation~\cite{chabanne2017privacy} --- are generally outperformed by Chebyshev interpolation, for non-linear functions involving operations such as $e^{-x}, \ln, \tanh$, as it is the case with complex AFs (c.f.,~\cite{miranda1996lecture} Table 5.2).

\changed{%Lastly, 
Piece-wise polynomial approximation can be parameterized by the degree of the polynomial ($\maxdegree$), the number of pieces ($\npieces$), and the Ring in which the MPC will be executed ($\ring$).  We establish a procedure to find a better tradeoff between accuracy loss and computational overhead.  %We  propose constraint optimization problem (COP), and 
%propose a new 
using 
application-specific heuristic that dynamically adjusts these parameters to find a desirable piece-wise polynomial.}
% $\solution = \langle \npieces, \maxdegree, \ring \rangle$ of AFs require finding an appropriate $\solution = \langle \npieces, \maxdegree, \ring \rangle$ % used for approximation that 
% balances performance overhead and accuracy loss, we treat this problem as a constraint optimization problem (COP), and propose a new application-specific searching heuristic  that dynamically adjust these parameters of $\solution$.} 
Specifically, we first pose this problem as a constraint optimization problem (COP) by setting a constraint of the maximum accuracy loss, say $\negligibleaccuracy$, that a practitioner can tolerate. Then we search for a $\langle \npieces', \maxdegree', \ring' \rangle$ that yields an approximation that has the lowest inference time under the constraint that accuracy loss is below $\negligibleaccuracy$.
\changed{We base our searching heuristic on simulated annealing (SA) which is a popular framework to solve COP.} 

Concretely, we start with an initial solution with high $\langle \npieces_{0}, \maxdegree_{0}, \ring_{0} \rangle$ so that it 
yields an approximation that has accuracy loss $\le \negligibleaccuracy$, but not necessarily the lowest inference time.
As a result the initial choice of  $\langle \npieces_{0}, \maxdegree_{0}, \ring_{0} \rangle$ may 
result in an approximation having pronounced inference time yielding an imbalance between performance and accuracy. To fix this, 
% Since lowering any of these parameters of $\langle \npieces_{0}, \maxdegree_{0}, \ring_{0} \rangle$  reduces performance overhead, to rectify this imbalance, we 
%\fixme{
we randomly make 
local adjustments to explore adjacent solutions of $\langle \npieces_{0}, \maxdegree_{0}, \ring_{0} \rangle$, and continue moving towards a solution $\langle \npieces', \maxdegree', \ring' \rangle$ under the SA framework 
that reduces performance overhead and accuracy loss remains less than $\negligibleaccuracy$ --- for a fixed number of iterations.
%} 

%to mitigate performance overhead while preserving minimal accuracy loss of $\negligibleaccuracy$, 
We carefully incorporate a number of application-specific techniques into the heuristic to avoid getting stuck on local optimal $\langle \npieces', \maxdegree', \ring' \rangle$.
For example, we find the approximation error threshold --- an important component of the heuristic --- via binary search instead of settling for a fixed value as prior work~\cite{NFGenFanCCS22}~(\secref{sec:selecting-threshold}). Furthermore, we introduce a DNN-specific modification to enhance the performance efficacy~(\secref{sec:designing-fcrude}).

We implement \sysname and perform extensive experiments using four different state-of-the-art DNN models with many hidden layers on diverse classification tasks. We find that $\sysname$ and $\nfgen$~\cite{NFGenFanCCS22} 
% (state-of-the approach to generate MPC-friendly approximation for non-linear functions) 
incur negligible accuracy loss compared to existing approaches~\cite{liu2017oblivious, li2022mpcformer, rathee2021sirnn}~(\secref{sec:accuracy}). 
Then, to compare performance overhead between \sysname and \nfgen, we incorporate their 
generated MPC-friendly approximation of complex AFs
to two state-of-the-art secure inference MPC libraries \abythree~\cite{aby3peter} and CryptFlow2~\cite{rathee2020cryptflow2}, and measure average inference time.
Our experiments reveal that our DNN model-specific optimizations make $\sysname$ 2$\times$--5$\times$ computationally more efficient 
% computationally efficient is kind of vague? repeat performance overhead. 
than $\nfgen$~\cite{NFGenFanCCS22} --- for DNN models having a high number of hidden layers, all while maintaining negligible accuracy loss. We have released $\sysname$ as an open source project~\cite{compact-github}.
% We are in the process of open-sourcing $\sysname$ with the next version of the paper.

% We provide a systematic way to approximate complex activation functions through a new importance-based Chebyshev approximation, and a method to find parameters that dynamically adjust between the accuracy and performance requirements. We use extensive empiricism to demonstrate the efficacy of our approach on a number of AFs and models, also demonstrating the generalizability of our approach. The contribution of the work is at par or exceeds recent works in the area of private inference using MPC that are published in top-tier conferences [18, 47, 43, 62]. 

\paragraph{Summary.} Our contributions are as follows: 
% \rcnote{Too many contributions need to  combine them into three contributions.}
\begin{newitemize}
    \item We present \sysname, a scheme that can generate MPC-friendly piece-wise polynomial approximations for 
    %popular 
    complex non-linear AFs. The generated approximation is generic and can be easily incorporated into state-of-the-art multi-party computation scenarios~(\secref{sec:overview}).

    \item The approximation technique used in our scheme is input density aware and accurately approximates regions with high input probability density while coarsely estimating regions with low input probability density~(\secref{sec:accurate-approximations}).
    
    \item \changed{We propose a  new searching heuristic based on simulated annealing framework to find parameters that dynamically adjust %and %\fixme{?} 
    % finds in 
    an approximation that have balance performance overhead and accuracy loss~(\secref{sec:time}).}
      
    \item We conduct extensive experiments and show that \sysname 
    generated MPC-friendly approximation of complex AFs have both negligible inference accuracy loss  than other DNN-specific approaches~\cite{liu2017oblivious, li2022mpcformer, rathee2021sirnn}, and 2$\times$--5$\times$ faster than $\nfgen$~\cite{NFGenFanCCS22}~(\secref{sec:experiments}). 
\end{newitemize}
% strategically
% }
% We believe that our proposed approach, \sysname, holds the promise for enhancing the efficiency and accuracy of secure inference in DNN models with complex AFs and contributes to advancement of secure multi-party computation techniques in real-world applications.

% \fixme{We believe....}

\section{Background and Related Work} \label{sec:background-related-work}
\def\figsize4{0.55}

\begin{figure*}
  % \centering 
  \footnotesize
  \pgfplotstableread[col sep = comma]{figures/data/silu.csv}\siludata
  \pgfplotstableread[col sep = comma]{figures/data/gelu.csv}\geludata
  \pgfplotstableread[col sep = comma]{figures/data/sigmoid.csv}\sigmoiddata
  \pgfplotstableread[col sep = comma]{figures/data/mish.csv}\mishdata
  \pgfplotstableread[col sep = comma]{figures/data/relu.csv}\reludata
  % \definecolor{color0}{rgb}{0.12156862745098,0.466666666666667,0.705882352941177}
  % \definecolor{color1}{rgb}{1,0.498039215686275,0.0549019607843137}
  % \definecolor{color2}{rgb}{0.172549019607843,0.627450980392157,0.172549019607843}
% \hpagessss{0.2}{0.24}{0.25}{0.2}{
% \centering
% \begin{subfigure}{0.15\columnwidth}
\begin{tikzpicture}[scale=\figsize4]
  \node[] at (3.5, 6.1) {\small $\relu(x) = \max(0, x)$};
      \begin{axis}[
          legend cell align={left},
          legend style={
            fill opacity=0.8, 
            draw opacity=1, 
            text opacity=1, 
            draw=white!80!black,
            at={(0.3,0.99)}
          },
          tick align=outside,
          tick pos=left,
          mark repeat=5,
          mark size = 2,
          x grid style={white!69.0196078431373!black},
          ymajorgrids=true,
          xlabel={\Huge $x$},
          xtick style={color=black},
          ylabel={\Huge $y$}, % $\fmpc(x)$}  %($\dv[2]{\fx(x)}{x})$},
          y grid style={white!69.0196078431373!black},
          ymajorgrids,
          ymin=-1, ymax=5,
          xmin=-5, xmax=5,
          ytick style={color=black}
      ]
      \addplot table[x={x}, y = {fx}] {\reludata};
      \addlegendentry{\LARGE $f(x)$}
      % \addplot table[x={x}, y = {fx1}] {\reludata};
      % \addlegendentry{\LARGE $f''(x)$}
  \end{axis}
  \end{tikzpicture}
  % {$\relu(x) = \max(0, x)$}
% \end{subfigure}
% \begin{subfigure}{0.25\columnwidth}
% \centering
  \begin{tikzpicture}[scale=\figsize4]
    \centering
    \node[] at (3.5, 6.1) {\small $\silu(x) = x/(1+e^{-x})$};
      \begin{axis}[
          legend cell align={left},
          legend style={
            fill opacity=0.8, 
            draw opacity=1, 
            text opacity=1, 
            draw=white!80!black,
            at={(0.3,0.99)}
          },
          tick align=outside,
          tick pos=left,
          mark repeat=10,
          mark size = 2,
          x grid style={white!69.0196078431373!black},
          ymajorgrids=true,
          xlabel={\Huge $x$},
          xtick style={color=black},
          ylabel={\Huge $y$}, % $\fmpc(x)$}  %($\dv[2]{\fx(x)}{x})$},
          y grid style={white!69.0196078431373!black},
          ymajorgrids,
          ymin=-1, ymax=5,
          xmin=-5, xmax=5,
          ytick style={color=black},
          % title=$\silu$,
      ]
      \addplot table[x={x}, y = {fx}] {\siludata};
      \addlegendentry{\Large $f(x)$}
      \addplot table[x={x}, y = {fx2}] {\siludata};
      \addlegendentry{\Large $f''(x)$}
  \end{axis}
  \end{tikzpicture}
  % {$\silu(x) = x/(1+e^{-x})$}
% \end{subfigure}
%   \begin{subfigure}{0.25\columnwidth}
  \centering
  \begin{tikzpicture}[scale=\figsize4]
    % \node[] at (3.5, 6.1) {\small $\gelu(x) = 0.5x\left(1 + \tanh\left[\sqrt{\frac{2}{\pi}} (x+0.044715x^3)\right]\right)$};
    \node[] at (3.5, 6.1) {\small $\gelu(x) \approx x/(1+e^{-1.702x})$$^\ddag$};
      \begin{axis}[
          legend cell align={left},
          legend style={
            fill opacity=0.8, 
            draw opacity=1, 
            text opacity=1, 
            draw=white!80!black,
            at={(0.3,0.99)}
          },
          tick align=outside,
          tick pos=left,
          mark repeat=10,
          mark size = 2,
          x grid style={white!69.0196078431373!black},
          ymajorgrids=true,
          xlabel={\Huge $x$},
          xtick style={color=black},
          ylabel={\Huge $y$}, % $\fmpc(x)$}  %($\dv[2]{\fx(x)}{x})$},
          y grid style={white!69.0196078431373!black},
          ymajorgrids,
          ymin=-1, ymax=5,
          xmin=-5, xmax=5,
          ytick style={color=black},
          % title=$\gelu$
      ]
      \addplot table[x={x}, y = {fx}] {\geludata};
      \addlegendentry{\Large $f(x)$}
      \addplot table[x={x}, y = {fx2}] {\geludata};
      \addlegendentry{\Large $f''(x)$}
  \end{axis}
  \end{tikzpicture}
  % \caption{$f(x) = \gelu(x) = x/(1+e^{-1.702x})$}
  % {$\gelu(x) = 0.5x\left(1 + \tanh\left[\sqrt{\frac{2}{\pi}} (x+0.044715x^3)\right]\right)$}
% \end{subfigure}
%   \begin{subfigure}{0.25\columnwidth}
  \centering
  \begin{tikzpicture}[scale=\figsize4]
    \node[] at (3.5, 6.1) {\small $\mish(x) = x\tanh(\ln(1 + e^x))$};
      \begin{axis}[
          legend cell align={left},
          legend style={
            fill opacity=0.8, 
            draw opacity=1, 
            text opacity=1, 
            draw=white!80!black,
            at={(0.3,0.99)}
          },
          tick align=outside,
          tick pos=left,
          mark repeat=10,
          mark size = 2,
          x grid style={white!69.0196078431373!black},
          ymajorgrids=true,
          xlabel={\Huge $x$},
          xtick style={color=black},
          ylabel={\Huge $y$}, % $\fmpc(x)$}  %($\dv[2]{\fx(x)}{x})$},
          y grid style={white!69.0196078431373!black},
          ymajorgrids,
          ymin=-1, ymax=5,
          xmin=-5, xmax=5,
          ytick style={color=black},
          % title=$\mish$
      ]      
      \addplot table[x={x}, y = {fx}] {\mishdata};
      \addlegendentry{\Large $f(x)$}
      \addplot table[x={x}, y = {fx2}] {\mishdata};
      \addlegendentry{\Large $f''(x)$}
  \end{axis}
  \end{tikzpicture}
  % {$\mish(x) = x\tanh(\ln(1 + e^x)) $}
% \end{subfigure}
\begin{flushleft}
  $^\dag$ Second derivatives $f''(x)$ equal to zero indicates that linear polynomials can easily approximate the function. 
  \\ $^\ddag$ More accurate version is $\gelu(x) = 0.5x(1 + \tanh[\sqrt{2/\pi} (x+0.044715x^3)]).$
\end{flushleft}
% \vspace{-1em}
  \caption{ 
    Complex activation functions (AFs) we focus in our work $f(x) \in \{\silu, \gelu, \mish\}$ and their second derivatives $f''(x)$. These AFs are hard to approximate accurately in regions close to zero where $f''(x) > 0$. We argue this is especially problematic for DNN models as the majority of the input to the complex AF falls to the region that are hard to approximate accurately (i.e., close to zero)  due to normalization (\figref{fig:normalization}). In contrast, $\relu(x)$ AF can be precisely  approximated with only two simple polynomials $\{\poly_1, \poly_2\}$ which are $\poly_1(x) =0 $ when $x < 0$  and $\poly_2(x) = x$ when $x \ge 0$.   
  }
  \label{fig:actfuncs}
\end{figure*}
   
This section summarizes relevant background and prior work from deep neural networks (\secref{sec:related-work-dnn}) and 
cryptographic techniques developed for solving the secure inference problem (\secref{sec:related-work-mpc}).  
% \vspace{-0.5cm} 

\subsection{Deep Neural Network Preliminaries} \label{sec:related-work-dnn}
\paragraph{Activation Functions (AFs).}
\label{sec:complex-activation-functions}
% importance of activation functions.
AFs are used for adding non-linearity to the learning process and play a major role 
in enhancing the training capabilities and accuracy of the DNN models. 
Many contemporary models use $\relu$ AF, which makes a hard gating decision based on the input sign~(\figref{fig:actfuncs}). Despite being theoretically simple, $\relu$ provides remarkably faster convergence and performance in practice~\cite{krizhevsky2017imagenet, nair2010rectified}. 
However, $\relu$ outputs a value of zero whenever the input is negative, and as such, the neural network loses a certain
amount of valid information as soon as inputs become negative. 
This drawback prompted ML communities to develop complex AFs, overcoming the limitations of $\relu$.

\paragraph{Complex AFs.}
% \fixme{Add a bit more details related to why ML communities are going to complex AF}
In recent years, a range of complex AFs, such as $\silu$~\cite{elfwing2018sigmoid}, 
$\gelu$~\cite{hendrycks2016gaussian}, and $\mish$~\cite{misra2019mish}, have emerged surpassing the performance of $\relu$ in state-of-the-art DNN models applied across
computer vision, natural language processing, and reinforcement learning applications. 
These AFs as shown in~\figref{fig:actfuncs}, 
are smooth and continuous, can handle small weight changes, and aid in effectively regularizing DNN models. For example, Hendrycks et al.~\cite{hendrycks2016gaussian} empirically illustrated the robustness of $\gelu$-trained DNN models
against noisy inputs, often surpassing the accuracy of $\relu$-trained models. 
Ramachandran et al.~\cite{ramachandran2017searching} used automatic search techniques to uncover $\silu$ (also called Swish). This complex AF improved image classification accuracy of Inception-ResNet-v2 model by 0.6\% and by 0.9\% of Mobile NASNET-A model by simply substituting it with $\relu$. Misra et al.~\cite{misra2019mish} proposed the self-regularized AF $\mish$ that exhibits superior performance compared to AFs for YOLOv4, ResNet models. 

Hence, complex AFs offer a compelling advantage in building better-performing models in terms of
convergence and classification accuracy when compared to $\relu$. Unfortunately, unlike $\relu$, which is relatively easy to compute for secure evaluation, these complex AFs exhibit a higher degree of non-linearity near the region close to zero as shown in~\figref{fig:actfuncs}. 
This makes their use with existing MPC techniques challenging. 
In this work, we address this limitation by designing an MPC-friendly version of these three complex AFs. We refer more interested readers to~\appref{app:related-work} for additional details on other complex AFs used in neural networks that lie outside the scope of this work.
\begin{figure}[t]
    \centering
    \includegraphics[width=\linewidth]{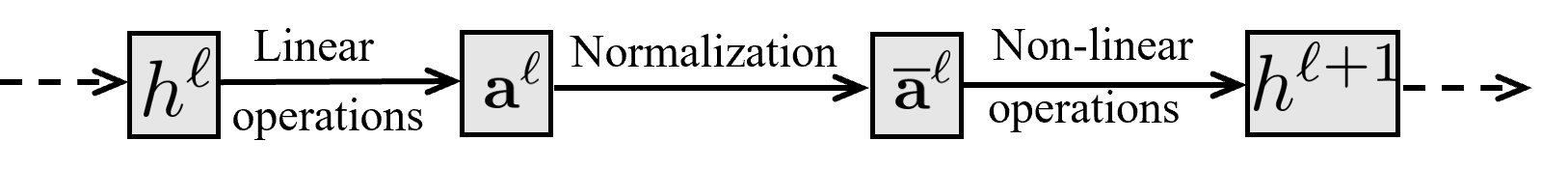}
    \caption{The output of the linear operations ($\mathbf{a}^\ell$) are normalized to $\overline{\mathbf{a}}^\ell$ using ~\eqnref{eqn:bn} before they are forwarded for applying non-linear operations involving complex activation functions (AFs). % \minote{Need to polish this Figure.}
    % show input x^\ell and output x^\ell + 1
    }
    \label{fig:normalization}
   
\end{figure}

\paragraph{Batch normalization.}  \label{sec:normalization}
Batch normalization (BN) is used to address \textit{internal covariance shift} problem in neural networks --- which happens when a layer's input distribution changes abruptly due to its dependency on previous layers~\cite{ioffe2015batch}. BN lends stability to the training process by reducing dependence on initial parameter selection, requiring a lower learning rate, and number of epochs. BN is performed on the outputs of the linear transformations, and normalized outputs are forwarded to non-linear AFs. 
Thus, non-linear AFs receive normalized inputs.
% Non-linear component applies non-linear activation functions on the normalized values without changing the input dimension.
% \rcnote{Bold face the vectors?}
\figref{fig:normalization} illustrates BN process 
for  $\ell^{\textsf{th}}$ layer where input to the linear operations is $h^\ell$ and output is 
$\mathbf{a}^\ell = {w^\ell}^Th^{l}$.
Assume $\mathbf{a}^\ell = \Big(a^\ell_1, a^\ell_2, \cdots, a^\ell_d\Big)$  is $d$-dimensional.
If the population mean  and variance are $\mathsf{E}[\mathbf{a}^{\ell}], \mathsf{Var[\mathbf{a}^{\ell}]}$ respectively,  
then $\mathbf{a}^\ell$ is normalized to $\overline{\mathbf{a}}^l$  using the following \eqnref{eqn:bn} such that the probability distribution of $\overline{a}^l$  follows a normal distribution with zero mean and unit variance:
% \inlineequation[eqn:bn]{  
\begin{equation}
\overline{\mathbf{a}}^{\ell}_{k} =  (\mathbf{a}^{\ell}_{k} - \mathsf{E}[\mathbf{a}^{\ell}_{k}])/\mathsf{Var[\mathbf{a}^{\ell}_{k}]}~\refstepcounter{equation}(\theequation)\label{eqn:bn}
%}
 %    \label{eqn:bn} 
\end{equation}

BN is widely used in state-of-the-art DNN models to calibrate the input to the  
non-linear AFs during both training and inference phases. 
This makes BN a good estimator of the input density to complex AFs in DNN models during inference. 
Our scheme leverages this estimation to improve the inference accuracy of the generated 
MPC-friendly approximations. 
% without increasing the performance overhead.
%affecting the performance adversely.

\subsection{Secure Inference for DNN models} \label{sec:related-work-mpc}
State-of-the-art MPC techniques enable computation on encrypted data and have been used to address the secure inference problem. Generally, a client encrypts their input and sends the encrypted input to a cloud service. The cloud service performs inference using trained DNN models over the encrypted input. Typically, MPC techniques are optimized for linear transformations (e.g., addition, matrix-vector multiplications, etc.). Therefore, computing non-linear operations involved in secure inference (e.g., non-linear AFs) is one of the main challenges. 

\paragraph{$\relu$ specific secure inference.} 
Given the popularity of $\relu$ in practical deployments of DNNs, 
recent research has mostly focused on the use of 
$\relu$. 
% \changed{
Early two works~\cite{chabanne2017privacy,gilad2016cryptonets} in this area generate Fully Homomorphic Encryption (FHE) friendly approximation of $\relu$ AF for secure inference.
% } 
Recent work focuses on 
MPC friendly approximation of $\relu$~\cite{wagh2020falcon,wagh2019securenn, rathee2020cryptflow2, chandran2021simc, huang2022cheetah}.

For example, Rathee et al.~\cite{rathee2020cryptflow2} propose a novel 2PC protocol
for secure comparison and division for efficient evaluation of $\relu$  
in semi-honest settings. 
Follow-up works 
extend this protocol to the malicious client threat model~\cite{chandran2021simc, huang2022cheetah}.
However, $\relu$ specific optimizations proposed in the aforementioned methods do not generalize to other complex AFs.
Another set of methods uses Garbled Circuits (GC) for secure evaluation 
of $\relu$
~\cite{mishra2020delphi, lehmkuhl2021muse, juvekar2018gazelle, riazi2018chameleon}. However, communication overhead limits its applicability to
shallow DNN models (less than seven layers). It is challenging to generalize these methods to wide DNN models that use complex AFs other than $\relu$ for 
secure inference.

% Another technique to handle the  
A different approach for computing non-linear AFs
efficiently in the encrypted domain is by  restricting 
the way DNN models are trained. 
For example, Riazi et al.~\cite{riazi2019xonn} leverage GC based protocol 
for secure 
inference on binary neural networks (BNN).
However, retraining and pruning proprietary models with these restrictions could be costly 
and oftentimes practically infeasible.   
Imposing such limitations on the training process can also impact the performance of DNNs in practice.
Pereteanu et al.~\cite{pereteanu2022split} introduce the notion of partially private DNN models
such that the middle part of the model is sent in plaintext to clients to reduce communication overhead. 
However, in practice,  cloud service providers would want to keep their full part of the DNN model 
secret lest revealing any part of the property model leaks sensitive information, resulting in severe business consequences.

In summary, while many promising works~\cite{ng2023sok} have focused on secure inference for $\relu$-based DNNs, our work focuses on novel complex AFs that have been shown to outperform $\relu$ 
and are getting traction in the ML community. 
%We refer the readers to ~\cite{ng2023sok} for a recent comprehensive survey for secure inference on ReLU-based.

\paragraph{Secure inference for other non-linear AFs.} 
A common approach for secure inference involving non-linear AFs is by 
approximating them with low-degree polynomials. These
polynomials are easy to compute for MPC frameworks and 
thus are MPC-friendly. The challenge is not
to degrade the inference accuracy, as the approximation error can cause incorrect results. 
Delphi~\cite{mishra2020delphi}, for example, runs a planner that balances which AF
can be replaced with low-degree polynomials without introducing too many inaccuracies and achieving a significant communication benefit. CryptoNet~\cite{gilad2016cryptonets}
CryptoDL~\cite{hesamifard2017cryptodl},  MiniONN~\cite{liu2017oblivious} also, use similar ideas for approximating non-linear AFs. 
However, they are application-specific, and switching to another 
application degrades accuracy significantly~\cite{garimella2021sisyphus} 
due to small errors getting propagated resulting in numeric instability. 
In addition, MiniONN~\cite{liu2017oblivious} is heavily focused on $\sigmoid$ AF --- which is essential for logistic regression models. 

However, as we will show in~\secref{sec:accuracy}, 
when we use their recipe for generating MPC-friendly approximation of the complex AFs that we focus on in this work, the inference accuracy decreases drastically.
% check with chatGPT (done once looks okay)
Recently, Fan et al.~\cite{NFGenFanCCS22} proposed $\nfgen$, a technique capable of converting popular non-linear functions --- used in scientific domains --- to MPC-friendly ones.
One may also choose to use this approximation-based approach to do the same for complex non-linear AFs. 
In fact, $\nfgen$ is the closest related work to ours. 
%as we also follow a similar approach---generating MPC-friendly approximations of the complex AFs using a number of piece-wise polynomials. 
However, 
$\nfgen$ is not specifically customized for widely used complex AFs inside DNN models.
Absence of such customized techniques makes $\nfgen$ computationally less efficient % while maintaining a similar or better accuracy 
when we compare it with our scheme through extensive experiments~(\secref{sec:performance}).

% \fixme{\textbf{Our work.}}

% Thus, when we conduct extensive experiments over  is specifically tailored for widely used complex AFs
% and thus provides better performance efficacy as the DNN models
% get deeper while maintaining a similar or better accuracy.

% Another way to solve this problem is by generating MPC-friendly approximations of the complex 
% \fixme{Need to bring $\nfgen$ here and highlight the changes}.

% \fixme{Add GPU and TEE based solutions.}
% https://arxiv.org/pdf/2011.05905.pdf
% cryptGPU, Phirana.

\section{Problem Overview \& Design Goals} \label{sec:problem-formulation} 
\newcommand{\nshares}{\ensuremath n}
\newcommand{\secretThreshold}{\ensuremath t}
\newcommand{\result}{\ensuremath z}
In this section, we first formulate the problem of secure inference and detail the threat model~(\secref{sec:problem-overview}). Then we describe the design goals we want to ensure while developing \sysname~(\secref{sec:requirements}). 
\subsection{Problem Overview} \label{sec:problem-overview}
\begin{figure}[t]
    \centering
    \footnotesize
    \includegraphics[width=\linewidth]{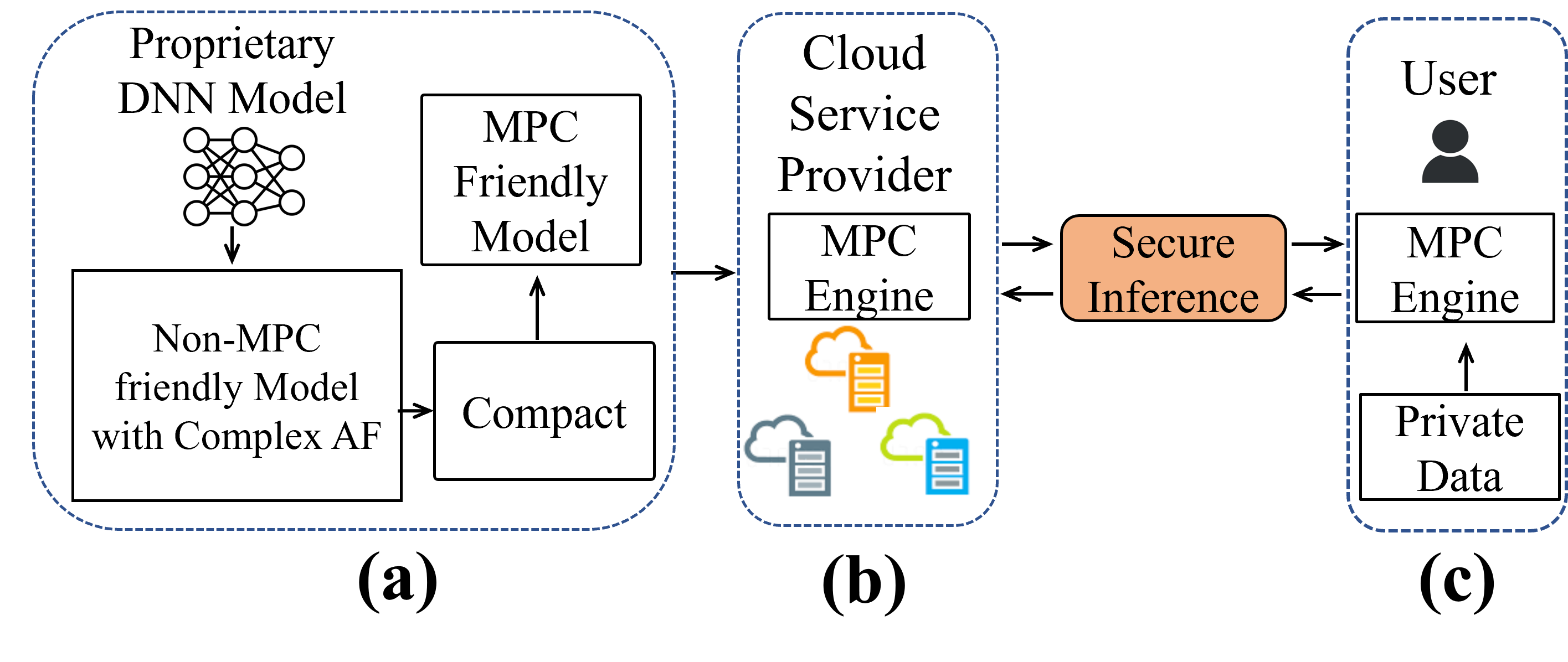}
    \caption{
        Secure inference in cloud-based deployment setting. 
        \textbf{(a)} Proprietary DNN model trained over private data that is not MPC-friendly due to complex non-linear activation functions (AFs) (e.g., $\silu, \gelu, \mish$). An MPC-friendly model is generated by replacing the complex AFs with their approximations using $\sysname$. \textbf{(b)} Next, we generate 
        $\nshares$ secret shares of the MPC-friendly model and distribute them with $\nshares$ computing servers (in this figure $\nshares = 3$) on the cloud. \textbf{(c)} To get the inference result, the client gets the private input data from the user, generates shares of it and distribute these with the  $\nshares$ servers. These servers on the cloud perform secure inference using an MPC engine and return the shares of the inference result to the client, and the client uses them to reconstruct the original inference result. 
    }
    \label{fig:overview}
    % \vspace{-0.5cm}
\end{figure}

% \subsection{Problem Formulation and Scenario Setup} \label{sec:problem-formulation-and-scenario-setup}
\paragraph{Problem formulation.}
We refer to the server holding the DNN model by $\modelowner$. 
The DNN model consists of $L$ layers, each comprising  linear transformations and non-linear complex activation function (AF) $\fx$. 
In between linear and non-linear complex AF, batch normalization is also present. 
We assume a machine learning as a service (MLasS)~\cite{forbes} inspired scenario where weights $\modelinput= [\weights_1, \weights_2, \cdots, \weights_L]$  of all $L$ layers of the model have already been trained, and the trained model is being used to provide cloud-based inference $z$ over client ($\client$)  uploaded input $\clientinput$  using~\eqnref{eqn:problem-formulation}.
% \inlineequation[eqn:problem-formulatio]{
\begin{equation}
\result := \fx(\weights_L \cdot \cdots \fx(\weights_2 \cdot \fx(\weights_1 \cdot \clientinput)))
% ~\refstepcounter{equation}(\theequation)
\label{eqn:problem-formulation}
% }.
\end{equation}
%~\eqnref{eqn:problem-formulation}.  
% \begin{equation}
    
%    \label{eqn:problem-formulation}
%\end{equation}
% \fixme{Here refer to~\figref{fig:overview}}
The problem secure inference tackles is how to compute the above equation %~\eqnref{eqn:problem-formulation} 
\emph{obliviously} to satisfy the privacy needs of both $\client$ and $\modelowner$. 
This requires designing a system such that $\client$ knows nothing about model weights $\modelinput$ and $\modelowner$ learns nothing about $\clientinput$; yet $\client$ can get the inference result $z$.
Moreover, we need to achieve this privacy need with both negligible accuracy loss and
reduced performance overhead. 
% We allow revealing L. 

% \paragraph{Difficulty in computing non-linear AFs.}
%Regardless of the number of $\server$ participating, for all state-of-the-art protocols, 

\paragraph{Scenario Setup.}
Secure multi-party computation (MPC) techniques enable a set of mutually distrusting parties to compute a function over their private inputs without revealing the inputs to other parties. 
Most MPC platforms use a version of \emph{secret sharing}~\cite{beimel2011secret}.
A $(\secretThreshold, \nshares)$-secret sharing scheme divides a
secret input $s$ into $\nshares$ shares, such that any $\secretThreshold-1$ of these shares reveal no information
about $s$, whereas any $t$ shares allow complete reconstruction of $s$.
Based on this primitive, we consider the following scenario to solve the secure inference problem using MPC.

In our setup, there are two parties: the first one is $\modelowner$ who owns the trained model with weights $\modelinput$. 
The second party is the user $\client$ who queries the model with their private data $\clientinput$. 
% This machine learning-based secure inference service is provided by $\nshares$ parties which we refer to as computing servers.
In a typical secure inference system, a set of semi-honest computing servers help compute the secure inference. These servers do not have any input of their own but facilitate the secure computation procedure.   Neither the model owner $\modelowner$ nor the client $\client$ trust these servers with their private inputs; however, they trust them to follow the protocol specified. 

Secure inference is performed in two phases. In the first phase,  $\modelowner$  locally
generates $\nshares$ secret shares of their private data $\modelinput$ as $\llbracket \modelinput_1, \modelinput_2, \cdots \modelinput_{\nshares} \rrbracket$, and  distributes them amongst  $\nshares$ computing servers over the network. 
After this phase, when the client $\client$  wants to query the secure inference service with their privacy-sensitive data $\clientinput$, they need to generate  $\nshares$ secret shares of it locally  as
$\llbracket \clientinput_1, \clientinput_2, \cdots, \clientinput_\nshares \rrbracket$, and send these shares to the $\nshares$  computing servers. Finally, these $\nshares$ computing servers engage in an MPC protocol to securely compute~\eqnref{eqn:problem-formulation}, and generate $\nshares$ secret shares of the result $\result$. At the end of the protocol, $\secretThreshold$ of these secret shares of $\result$ are sent to the querying user who can combine these $\nshares$ shares to construct the final result $\result$.
Following prior work we consider two settings based on the number of computing servers: 
(i) $\nshares = 2$~\cite{mishra2020delphi,sirnn-lib}, and (ii) $\nshares = 3$~\cite{wagh2020falcon,wagh2020falcon,aby3}. 
In this study, we refer to them as 2PC and 3PC scenarios respectively.

% and $\client$ have to generate $\nshares$ secret shares of their private input data 
% $\modelinput$, and $\clientinput$ respectively

% When there are only two parties, we refer to as two-party computation. In our setting we will focus on the two-party scenario.  
% % MPC techniques provide a generic solution when
% Formally, let two parties hold private inputs $X$ and $Y$, and they want to securely compute a function $Z = \mathcal{F}({X},{Y})$ without revealing their private inputs to each other.  \rcnote{Add something about these $n$ non-colluding computing servers here.}
%  Based on this,
% each party generates $\nshares$ secret shares of their private data ${X}$ and ${Y}$ locally and distribute them amongst  $\nshares$ non-colluding servers. 
% }\rcnote{Unclear how many parties are there in our setting? Where does n-server comes? Is it only 2-party?}

% To solve the secure inference problem using MPC, , and share them with $\nshares$ non-colluding  servers. At the end of the protocol $\client$ gets $\secretThreshold$ secret shares of inference result to reconstruct $z$.

% \changed{
\paragraph{Threat model and scope.}
% Different MPC protocols can protect data privacy against different security assumptions regarding the adversarial power $\adv$ of these  $\nshares$ servers.
% In this work, we assume a general setup of an MPC protocol and henceforth inherit their security requirements. 
% More specifically, we assume these 
% the adversary $\adv$ is parameterized by four dimensions~\cite{lindell2020secure}. 
% They are i) corruption strategy (static, adaptive, proactive), ii) type of $\adv$ in terms of how they are following the protocol (semi-honest, malicious, covert), iii) corruption ability $\adv$ has (honest, dishonest majority), and iv) power of the $\adv$ (informational, computational-secure). 
In this work, the techniques we use do not apply any restriction on how the adversary $\adv$ is modeled by the MPC scheme or the type of secret sharing being used by the MPC scheme. 
Henceforth we can inherit the security requirements of the underlying MPC scheme. 

That being said, the majority of the existing works on secure inference assume that these $\nshares$ computing servers % and users 
are semi-honest
 % honest-but-curious 
 (i.e., adversaries who do not deviate from the protocol but try to learn as much information as possible as from the messages they receive)~\cite{ng2023sok}.
In practice, this can be achieved by placing these computing servers under the regulation of a trusted organization, and monitoring that they are following the MPC protocol. In our experimental evaluation, we adopt this threat model.
% from two secure inference MPC libraries \abythree~\cite{aby3peter}, and CryptFlow2~\cite{rathee2020cryptflow2}. 

We note that 
our scheme  does not guarantee protection of $\modelinput$, and $\clientinput$ against attacks such as training data poisoning~\cite{shafahi2018poison}, model inversion~\cite{fredrikson2015model}, adversarial examples~\cite{goodfellow2014explaining}, membership inference attacks~\cite{carlini2022membership}, etc. One wishes to do so should employ defenses from existing literature, and whether MPC schemes can be leveraged to provide protections against such attacks is an open question as discussed further in~\secref{sec:discussion}.

% For example, consider the scenario in the medical domain, as shown in~\figref{fig:overview}, 
% where a DNN model $\modelinput$ has been trained by a trusted organization (e.g., NIH\footnote{\url{https://www.nih.gov/}}) leveraging substantial computational resources and exclusive access to users' private health records. To preserve the privacy of the proprietary DNN model, NIH can generate secret shares of the model and distribute them across
% $\nshares$  non-colluding servers, possibly hosted by different hospitals and regulated by NIH.
% When patients submit their private health data $\clientinput$, they can generate $\nshares$ secret shares  
% $\clientinput = \llbracket \clientinput_1, \clientinput_2, \cdots, \clientinput_\nshares \rrbracket$
% and share them with the $\nshares$ different hospitals. In this way, the patient learns the final result without learning anything about the model or revealing their private information to any hospital.

\paragraph{Motivating example.} 
One of the motivating realizations of secure inference scenarios can be in the medical domain as pictured in~\figref{fig:overview}. 
In particular, where a DNN model weights $\modelinput$ has been trained by a trusted organization (e.g., National Institutes of Health) leveraging substantial computational resources and exclusive access to users' private health records.
% \footnote{\url{https://www.nih.gov/}} 
To preserve the privacy of the proprietary DNN model, NIH can generate secret shares of the model $\modelinput$ % $\modelinput = \llbracket \modelinput_1, \modelinput_2, \cdots \modelinput_{\nshares} \rrbracket$ 
and distribute them across
$\nshares$ different 
semi-honest
% honest-but-curious 
computing servers, possibly hosted by different hospitals.
When patients submit their private health data $\clientinput$, they can generate $\nshares$ secret shares  
%$\clientinput = \llbracket \clientinput_1, \clientinput_2, \cdots, \clientinput_\nshares \rrbracket$
and share them with the $\nshares$ different hospitals. In this way, the patient learns the final result without learning anything about $\modelinput$  or revealing their private information $\clientinput$ to any hospital. 

\paragraph{Difficulty in computing non-linear AFs.}
A major bottleneck while running the MPC protocol is computing $\fx(x)$ securely shown in~\eqnref{eqn:problem-formulation}. 
This is because $\fx(x)$ is non-linear, which consumes most of the 
communication and latency costs of the overall protocol execution, as illustrated by many prior works
~(e.g., Rathee et al.~\cite[Table 6]{rathee2020cryptflow2}).
Linear operations (i.e., matrix-vector multiplication) are  
less expensive comparatively.

\newcommand{\goal}[1]{{\circled{#1}}}
\newcolumntype{T}[3]{%
>{\adjustbox{angle=#1, left, minipage=#3, lap=\width-(#2)}\bgroup}%
l%
<{\egroup}%
}
\newcolumntype{P}[1]{>{\centering\arraybackslash}p{#1}}
\newcommand*\rot{\multicolumn{1}{T{50}{4em}{1.8cm}}}% no optional argument here, please!
\begin{table}[t]
    \centering
   \gamesfontsize
    \begin{tabular}{l*{4}{P{1.2cm}}}
    \toprule
        \multirow{2}{0.8cm}{Method} & \multirow{2}{1.2cm}{\goal{1} Supp. cmplx. AF} & \multirow{2}{1.2cm}{\goal{2} Supp. many HL} & \multirow{2}{1.2cm}{\goal{3} Comp. w/ MPC libs} & \multirow{2}{1.8cm}{\goal{4} Supp. any training proc.} \\ \\ \midrule
        CryptFlow2~\cite{rathee2020cryptflow2}$^\bigstar$ & \emptycirc & \fullcirc & \emptycirc & \fullcirc \\
        SIMC~\cite{chandran2021simc}$^\bigstar$ & \halfcirc & \fullcirc & \emptycirc & \fullcirc \\
        Cheeta~\cite{huang2022cheetah}$^\bigstar$ & \emptycirc & \fullcirc & \emptycirc & \fullcirc \\
        Delphi~\cite{mishra2020delphi}$^\bigstar$$^\mathsection$ & \emptycirc & \fullcirc & \emptycirc & \emptycirc \\
        % GAZELLE~\cite{juvekar2018gazelle} & \fixme{\halfcirc} & \fullcirc & \fullcirc & \fullcirc \\
        XONN~\cite{riazi2019xonn}$^\bigstar$ & \halfcirc & \fullcirc & \emptycirc & \emptycirc \\
        % Chameleon~\cite{riazi2018chameleon} & \halfcirc & \fullcirc & \fullcirc & \fullcirc \\
        SIRNN~\cite{rathee2021sirnn} & \halfcirc & \emptycirc & \emptycirc & \fullcirc \\
        % SecureML~\cite{mohassel2017secureml} & \halfcirc & \fullcirc & \fullcirc & \fullcirc \\
        MUSE~\cite{lehmkuhl2021muse}$^\bigstar$ & \halfcirc & \fullcirc & \emptycirc & \fullcirc \\
        SecureNN~\cite{wagh2019securenn}$^\bigstar$ & \emptycirc & \fullcirc & \emptycirc & \fullcirc \\
        FALCON~\cite{wagh2020falcon}$^\bigstar$ & \emptycirc & \fullcirc & \emptycirc & \fullcirc \\
        $\nfgen$~\cite{NFGenFanCCS22} & \fullcirc & \emptycirc & \fullcirc & \fullcirc \\
        \rowcolor{electriclime} \sysname (this) & \fullcirc & \fullcirc & \fullcirc & \fullcirc \\
        \bottomrule
        \\
        \multicolumn{5}{c}{ \fullcirc ~Yes; \mytab \halfcirc ~Unclear;  \mytab \emptycirc ~No} \\
    \end{tabular}
    \begin{flushleft}  
         $^\bigstar$ Support for large number of hidden layers (HLs) for these methods is experimentally validated for DNN models for $\relu$. \\ 
         $^\mathsection$ Although Delphi uses GC to evaluate non-linear layers, the MPC-friendly square function used to replace $\relu$ is specific towards $\relu$. Therefore, we do not consider it compatible with complex AFs.
    \end{flushleft}
    \caption{Comparison of related work with \sysname.} % (mainly 2019 onwards) on secure inference
    \label{tab:comparison}
    \vspace{-0.5cm}
\end{table}

\subsection{Design Goals} \label{sec:requirements}
% \paragraph{Design Goals.} 
While designing \sysname, we want to ensure DNN model designers are not restricted to the set of AFs and model architectures that MPC platforms support. We distill four criteria for this and show how prior work on secure inference
%, mainly from 2019 onwards, 
fail to satisfy one or more of these design goals in~\tabref{tab:comparison}.

\paragraph{\goal{1} Support complex AF.} We want our scheme to be compatible with the majority of the DNN models used by inference services. Therefore, in this work, we do not use $\relu$-specific optimizations.
%\sysname to  beyond using $\relu$ --- one of the simple AFs --- specific optimizations.
% This separates us from the majority of 
Majority of prior works are devoted to optimize $\relu$ and fail to satisfy this design goal~\cite{rathee2020cryptflow2,huang2022cheetah,mishra2020delphi,wagh2019securenn,wagh2020falcon}. Few works rely on garbled circuits (GC) to evaluate AFs, but experimental evaluations are limited to $\relu$ AF~\cite{chandran2021simc,lehmkuhl2021muse,riazi2019xonn,rathee2021sirnn}.
Therefore, it is unclear if these GC-based protocols can generalize to other complex AFs such as $\silu$, $\gelu$, and $\mish$. We marked them as \emph{unclear} in the first column of~\tabref{tab:comparison}. 

\sparagraph{\goal{2} Supports large number of hidden layers.} 
The error introduced due to replacing $\fx$ with its MPC-friendly approximation $\fmpc$ in~\eqnref{eqn:problem-formulation} can accumulate and possibly lead to a significant loss in accuracy as the number of hidden layers increases. 
Unfortunately, few prior works~\cite {rathee2021sirnn, liu2017oblivious, li2022mpcformer} that support complex AFs show significant accuracy loss for DNN models with a high number of 
hidden layers. $\nfgen$, however, does not exhibit this accuracy loss as the number of hidden layers increases, but this negligible accuracy loss comes at the cost of paying high-performance overhead.
We want our scheme to endure such accuracy loss as the number of hidden layers increases without increasing the performance overhead significantly.

\paragraph{\goal{3} Compatible with MPC libraries.}  
The secure inference procedure we develop should not only support a wide variety of AFs but also should be easy to implement. Implementations that require new cryptographic primitives for secure inference will be hard and slow to deploy.  
Therefore, in this work, we aim to design a scheme that can be implemented with generic
%is generic to 
MPC libraries currently in use.
Our solution only requires 
%, the MPC library only needs to support 
secure addition, multiplication, and comparison operations. 
This would also allow seamless transitioning from inference service using $\relu$ based DNN models to complex AF-based DNNs. Prior works other than $\nfgen$ do not satisfy this design goal. 
% support secure inference over DNN models. By doing so, we can also inherit all security guarantees provided to the existing MPC libraries seamlessly. 

% \item 
\paragraph{\goal{4} No restriction on training.} To handle accuracy loss with an increasing number of hidden layers, some prior works change the way DNN models are traditionally trained. For example, XONN requires restricting the weights of the DNN model to binary values (i.e., $\pm 1$); similarly, Delphi replaces certain AFs (i.e., $\relu$) with a square function during training. We believe this type of restriction poses additional constraints 
for deployment of secure inference 
as % already trained 
existing DNN models are most likely trained % traditionally 
without these restrictions, and % further attempts to adjust 
the weight of the already trained DNN models must be adjusted (e.g., fine-tuning % the trained DNN model 
by applying these restrictions) to comply with these protocols. 
% would be expensive. 
Therefore, we aim to design \sysname without any restriction on the training process of the DNN models.  
 
% \item \textbf{R2-negligible accuracy loss with more hidden layers:} 
% \item 
  
% \end{newitemize}

In summary, recent proposal in secure inference literature holds promise 
toward realizing secure inference; but they
do not % focus on 
satisfy the above-mentioned generality, deployability, and scalability aspects important for realizing secure
inference in practice.
% the real world.
We aim to bridge this gap via our designed scheme \sysname.

\begin{table}[t]
    \centering
    % \tabfontsize
    \begin{tabular}{@{}lp{7cm}@{}}
    \toprule
    Symbol & Description of the symbol \\ \midrule
    $\fx(x)$ & complex activation function we want to approximate.\\
    $\fmpc(x)$ &  MPC friendly piece-wise polynomial approximation of $\fx(x)$.\\
    $\Error$ & distance metric used to estimate the approximation error  between $\fx(x)$ and $\fmpc(x)$\\
    $\ethreshold$ & maximum threshold for approximation error.\\
    $\npieces$ & \# of piece-wise polynomials used for approximation.\\
    $\maxdegree$ & maximum degree of each of $\npieces$ piece-wise polynomials. \\
    $\ring_{\resolution,\decimal}$ & ring of size $\resolution$ is used in MPC library with last $\decimal$ bits representing the fractional parts. \\
    $\poly_i(x)$ & single polynomial approximating $\fx(x)$ between $[x_i, x_{i+1}]$ \\ 
    $[s,e]$ & the interval over which we are trying to approximate $\fx$ \\ 
    $[\alpha,\beta]$ & a continuous closed interval between $\alpha$ and $\beta$ \\ 
    $\density$ & probability distribution of the input to the activation function.\\ 
    \bottomrule
    \end{tabular}
    \caption{Notations used in this paper.}
    \label{tbl:notations}
    \vspace{-1.8em}
\end{table}
\newcommand{\fmpcbest}{\fmpc^{\mathsf{best}}}
\newcommand{\ethresholdlo}{\ethreshold_{\mathsf{lo}}}
\newcommand{\ethresholdhi}{\ethreshold_{\mathsf{high}}}
\newcommand{\ethresholdmid}{\ethreshold_{\mathsf{mid}}}
\newcommand{\thetacur}{\theta_{\mathsf{cur}}}
\newcommand{\trainingDataset}{\ensuremath D_{\mathsf{train}}}

\begin{figure*}[t]
    \centering
    % \tabfontsize 
    \hfpagess{0.4}{0.45}{
      \comments{Global parameters used by \sysname}\\ 
    $\fx \gets$ activation function to approximate between $[s, e]$\\  
    %$\Error \gets \MeanApproxError$ \comments{as defined in~\eqref{eqn:mean_approximation_error}} \\ 
    $\density \gets \normal(0, 1); \; \Error \gets \MeanApproxError; \; s \gets -5 ; e \gets 5 $ \\
    $\plainTextAcc \gets $ plaintext inference accuracy using $\fx$ \\
    $\negligibleaccuracy \gets 10^{-2}$ \comments{accuracy loss practitioners can tolerate.} \\ [4pt]
    %%%%%%%%%%%%%%%%%%%%%%%%%%%%%%%%%%%%%%%%%%%%%%%%%
    \underline{$\candidatelistgenerationalgo\left( 
      \theta    
    \right)$:}\\[2pt]
   % \textbf{Input:}  activation function \\
   % \textbf{Output:} \\ 
    $\langle \npieces, \maxdegree, \ring \rangle \gets \theta$ \\
    $\stepsize \gets (e-s)/\npieces$ \\
    $\fmpcbest \gets \emptyset;$ \;\; $\ethreshold_{\mathsf{lo}} \gets 0;$\;\; $\ethreshold_{\mathsf{hi}} \gets \mathsf{MAX\_APPROX\_ERROR}$ \\ [2pt]
    \textbf{while} $\ethreshold_{\mathsf{lo}} < \ethreshold_{\mathsf{hi}}$ \\
    \mytab $\fmpc \gets \emptyset;\;\; \ethresholdmid \gets (\ethresholdlo +  \ethresholdhi)/2; \;\; \alpha \gets s;$\;\; $\beta \gets \alpha + \stepsize$ \\ [2pt]
    % \mytab \mytab $\fmpc \gets \varnothing$ \\
    \mytab \textbf{while} $\beta < e$: \\
    \mytab \mytab $\poly \gets \interpolate \left(\fx, \maxdegree, \alpha, \beta\right)$ \\ 
    \mytab \mytab $\ecrude  \gets \Error\left(\density, \fx, \poly, \alpha, \beta\right)$ \\ 
    \mytab \mytab  \textbf{if} $\ecrude > \ethresholdmid/\npieces$: \\
    \mytab \mytab \mytab $\fmpc \gets \fmpc \bigcup \{\poly\} ;\;\; \alpha \gets \beta$  \\
    \mytab \mytab $\beta \gets \beta + \stepsize $ \\
    % \mytab \textbf{if} $|\fmpc| \le  m:$ \\
    \mytab $\secureInfAcc \gets $ compute accuracy using $\fmpc$ on $\ring$ \\
    \mytab \textbf{if} $(\plainTextAcc - \secureInfAcc)/\plainTextAcc \le \negligibleaccuracy \wedge |\fmpc| \le  m:$ \\ 
    \mytab \mytab $\fmpcbest \gets \fmpc$; \;\; $\ethreshold_{\mathsf{hi}} \gets \ethreshold_{\mathsf{mid}}$ \\[2pt] 
    \mytab \textbf{else:} \; \; %\\  \mytab \mytab 
    $\ethreshold_{\mathsf{lo}} \gets \ethreshold_{\mathsf{mid}}$ \\ [2pt]
    $\ret~\fmpcbest$
    }
    {
    % \nonl \underline{$\ApproximationAlgo$} $\Big (\fx, \density = \normal(0, 1), \Error = \MeanApproxError, s = -10, e = 10 \Big )$: \\ [2pt]
    
    %%%%%%%%%%%%%%%%%%%%%%%%%%%%%%%%%%%%%%%%%%%%%%
    \underline{$\ApproximationAlgo()$:} \\ [2pt] 
    $\thetacur \gets \langle \npieces_0,\maxdegree_0,\ring_0\rangle $     \comments{Initial solution} \\
    $\fcur \gets \candidatelistgenerationalgo(\thetacur)$ \\ 
    \comments{Solving the constraint optimization problem in~\eqnref{eqn:optimization}} \\ 
    \textbf{for} $\step\get 1$ to $\step_{max}$ \comments{$\step_{max} \gets 10$ for our experiments}\\ 
    \mytab $T_\step \gets \chi_0/\log(1+\step)$  \comments{$\chi_0 \gets 0.2$ for our experiments}\\ 
    \mytab $\theta_\step \gets \mathsf{GenerateNeighbour}(\tcur)$ \\
    \mytab $\fmpc^\step \gets \candidatelistgenerationalgo(\theta_\step)$ \\
    \mytab \textbf{if} $\fmpc^\step = \emptyset$: \mytab \textbf{continue} 
    \comments{cannot find an $\fmpc$ with $(\plainTextAcc - \secureInfAcc)/\plainTextAcc \le \negligibleaccuracy$}\\ 
    \mytab $r \getsr U_{[0, 1]}$ \\ 
    \mytab \textbf{if} $\exp\left((\Time(\fcur)-\Time(\fmpc^\step))/T_\step\right) > \randomNumber$: 
    % \comments{\eqnref{eqn:update-cur}} 
    \\ 
    \mytab \mytab $\fcur \gets \fmpc^\step; \mytab \tcur \gets \theta_{\step}$ \\[2pt]
    $\ret~\fcur$\\[6pt]
    \medskip 
    \underline{$\generateNeighbour(\theta)$:}\\
    $\langle \npieces, \maxdegree, \ring \rangle \gets \theta$ \\
    Sample $z_1,z_2 \in \mathbb{Z}$ according to the probability density function $p(z) = 1/3 \cdot 2^{-|z|}$\\ 
    $\npieces' \gets \npieces + z_1$; \;\; $\maxdegree' \gets \maxdegree + z_2$ \\ 
    $\resolution' \getsr \{128, 84, 64, 32\};  \gamma_2 \getsr \{1.5, 2, 2.5,3, 3.5, 4\}$\\
    $\decimal' \gets \floor{\resolution'/\gamma_2}$\\
    $\theta' \gets \langle \npieces', \maxdegree', \ring'_{\langle \resolution',\decimal' \rangle} \rangle $  \\ 
    $\ret~\theta'$ 
    }
    % \begin{flushleft}
    %   \tabfontsize  
    % $^\ddagger$ \\ 
    % $^\dagger$$\candidatelistgenerationalgo$ is described in~\secref{sec:accurate-approximations}, and 
    % % $\ApproximationAlgo$ and $\generateNeighbour$ are described in~\secref{sec:time}.  % \rcnote{Use threshold instead of $\sim$}
    % \end{flushleft}
      \caption{
      \textbf{(Right-top)} $\ApproximationAlgo$ procedure to find 
    an MPC-friendly approximation $\fmpc$ of the complex activation function (AF) $\fx$~(\secref{sec:time}). The procedure balances the trade-off between inference accuracy loss and performance overhead using an application-specific optimization approach (simulated annealing).
    % During our experiments we set $\chi_0 = 0.2$ and $k_{max} = 10$. 
    It uses two sub-procedures---$\generateNeighbour$ to generate a random neighbor $\theta'$ from a given $\theta$ (shown Right-bottom) and  
      $\candidatelistgenerationalgo$ to approximate the region $[s, e]$ accurately using a set of at most $\npieces$ polynomials (shown \textbf{Left}) with degree $\leq \maxdegree$ (\secref{sec:accurate-approximations}). Notations are explained briefly in~\tabref{tbl:notations}.
    }
      \label{algo:candidateListGen}
\end{figure*}

%%% Local Variables:
%%% mode: latex
%%% TeX-master: "../main"
%%% End:

\section{Design of \sysname} \label{sec:our-piece-wise-poly}
In this section, we first give an overview of our scheme in~\secref{sec:overview}. Then, we gradually detail  
our scheme in~\secref{sec:accurate-approximations} and~\secref{sec:time}. We sketch our scheme in~\figref{algo:candidateListGen} with a summary of used notations in~\tabref{tbl:notations}.

\subsection{Overview of $\sysname$} \label{sec:overview}
\paragraph{Piece-wise polynomial approximation approach.} Our scheme $\sysname$ follows the idea of approximating a complex activation function (AF) using a number of piece-wise polynomials.
First, we observe that complex AF can be approximated easily using linear functions outside a certain range. (Fan et al.~\cite{NFGenFanCCS22} made similar observations for $\sigmoid$.) 
Therefore, we only need to approximate  a small range of $x$ values, say $[s, e]$. 
%, which require approximation.
We will approximate $\fx$ using a piece-wise polynomial function, with $\npieces$ pieces $[\poly_1, \poly_2, \cdots, \poly_m]$ defined as follows: 
\begin{equation}
    \fmpc(x) = \sum_{i=0}^{m+1} \indicator_i(x)\cdot \poly_i(x)
    \label{eqn:approx}
\end{equation}
where $\indicator_i(x) = 1$ if $x\in (x_{i-1}, x_i]$, and $0$ otherwise, for all $i \in \{0, 1, \ldots, m+1\}$, $x_{-1} = -\infty$, $x_0 = s$, $x_m = e$, and $I_{m+1}(x) = 1$, if $x>e$, and 0 otherwise.
% what is f_1 and f_{m+1}.
The functions $\indicator_i$ define the pieces, and functions $\poly_i$ define the polynomials.
\changed{From~\figref{fig:actfuncs}, it is easy to see that when $x \le -5$ or  $x \ge 5$, $\fx(x)$ becomes equal (or very close) to zero and $x$, respectively. 
As such,  we can set $\poly_0(x) = 0$, and $\poly_{m+1}(x) = x$ by maintaining  $s \le -5$, and $e \ge 5$
for all complex AFs.}
For the other polynomials, we impose an additional constrain that $\poly_{i}$ % $\deg(\poly_i)$ 
must be of degree  $\maxdegree$ or less,  $\forall i \ \deg(\poly_i) \le \maxdegree$ as following. 
%shown in~\eqnref{eqn:polynomial}
% \bnm
\begin{equation}
    \poly_i(x) = a_0 + a_1x + a_2x^2 + \cdots +   a_{k}x^k
    \label{eqn:polynomial}
% \enm
\end{equation}
The above-mentioned approach is not specific towards $\relu$  and can approximate complex AFs (e.g., $\silu$, $\gelu$, $\mish$) --- satisfying design goal \goal{1} as described in~\secref{sec:requirements}. Furthermore~\eqnref{eqn:approx}, and (\ref{eqn:polynomial}) comprise of three math operations $\addmathoperation$, $\mulmathoperation$, and $\comparemathoperation$ and the majority of the MPC libraries support these three math operations and thus \sysname generated approximations of AF 
% ---making $\sysname$ generic to the MPC library being used and thus 
satisfy design goal \goal{3}.  % and easy to extend to malicious security. 

However, generally approximation-based approaches tend to be inaccurate~\cite{kelkar2022secure}. 
% recent work~\cite{kelkar2022secure} has highlighted one limitation of adopting such
% or introduce much computational overhead. % in the process of making them accurate. 
Thus, maintaining negligible accuracy loss with increasing hidden layers (design goal \goal{2}) and imposing no restriction on training (design goal \goal{4}) at the same time become challenging.  
We address this challenge by developing the techniques described in the subsequent sections.  

\subsection{Generating Accurate Approximations} \label{sec:accurate-approximations}
To approximate $\fx$  for a given region $[s,e]$ using  $\npieces$  piece-wise polynomials with degree at most $\maxdegree$ and has negligible accuracy loss, we use an opportunistic approach $\candidatelistgenerationalgo$ as shown in~\figref{algo:candidateListGen}.
We use an interpolation technique similar to the one proposed in $\nfgen$~\cite{NFGenFanCCS22}.  
%Note that our approach, as mentioned earlier, shares similarities with $\nfgen$~\cite{NFGenFanCCS22} only on the interpolation aspect (described in~\secref{sec:interpolation} for completeness). 
However we use dynamic  approximation which 
%Other techniques, described next, are tailored to the complex AFs used in DNN models and unique to % $\sysname$. 
% This 
makes \sysname computationally more efficient than $\nfgen$ as the number of hidden layers increases (experimentally illustrated in~\secref{sec:performance}). % a design goal NFGEn fails to achieve. 
% We now describe these techniques employed by \sysname in detail as follows.
\subsubsection{Computing $\density(x)$} \label{sec:density}

We aim to design approximate polynomials that are close to accurate on likely values of $x$, meaning higher probability according to $\density(x)$, while may have higher error on values of $x$, which are less likely. A challenge, however, is how to estimate the distribution $\density$.  
Interestingly, in DNN models, the inputs to an AF are first batch normalized (BN) using~\eqnref{eqn:bn}, to help the network converge faster during training (discussed earlier in~\secref{sec:normalization}).
Therefore, the set of values the AF is computed on is distributed (approximately) as a normal distribution with zero mean and unit standard deviation.
The approximating piecewise polynomial, therefore, should ensure low error on highly likely inputs, whereas on low probable inputs, it may allow making a higher error.

Our key insight is that $\density(x)$ can guide us to focus 
on approximating those regions more accurately 
where $\density(x)$  is high. 
% In contrast, we can get away with 
% a less accurate approximation where $\density(x)$
% is close to zero without degrading the accuracy significantly. 
% Moreover, due to BN applied on DNN layers prior to applying the AFs, 
We estimate $\density(x)$ using a standard normal distribution $\normal(0, 1)$, and 
% However, incorporating $\density(x)$ to the approximation procedure is not straightforward, and we 
use a customized function to compute the approximation error that takes into account $\density(x)$, as we describe next.
We remark on one caveat of this design choice: \sysname becomes reliant on BN as we discuss further in~\secref{sec:discussion}.

% \vspace{-0.85cm}
\subsubsection{Designing $\Error$} \label{sec:error} 
To incorporate $\density(x)$ to the approximation procedure, we customize an
approximation error  which we refer to as \textit{weighted mean approximation error} denoted by $\MeanApproxError$. 
\begin{multline}
    \MeanApproxError(\density, \fx, \poly, \alpha, \beta) = \\
    \frac{1}{(\beta-\alpha)} \int \limits_{\alpha}^{\beta}\density(x) \cdot |\fx(x) - \poly(x)| dx
    \label{eqn:mean_approximation_error}
\end{multline}
As shown in~\eqnref{eqn:mean_approximation_error},  in addition to considering how accurately $\poly(x)$ estimates $\fx(x)$ for a given region between $\alpha$ and $\beta$, $\MeanApproxError$   also takes $\density(x)$ into account.
$\nfgen$~\cite{NFGenFanCCS22} uses \textit{max approximation error}, which we denote by $\MaxApproxError$ as a way to design $\Error$. 
\bnm    \MaxApproxError(\fx, \poly, \alpha, \beta) = \max \limits_{x \in [\alpha, \beta]} |\fx(x) - \poly(x)|
    \label{eqn:max_approximation_error}
\enm
We choose to use $\MeanApproxError$ over $\MaxApproxError$ as it is easy
to guide the approximation process via $\density(x)$ using $\MeanApproxError$.  

\subsubsection{Selecting a threshold $\ethreshold$ for approximation error $\MeanApproxError$}
\label{sec:selecting-threshold}
A straightforward ad hoc way to ensure the accuracy of the approximation is to set a fixed approximation error threshold  ($\ethreshold$) and consider an approximation accurate if approximation error calculated via $\Error$ is $\le \ethreshold$. 
$\nfgen$ follows this ad hoc approach and sets $\ethreshold = 10^{-3}$. 
Via empirical experimentation, they observed that if  $\MaxApproxError \le 10^{-3}$, then the generated approximation, when used in logistic regression and $\chi^2$ testing, does not degrade accuracy without adding much performance overhead. 

We refrain from setting a fixed $\ethreshold$ for \sysname as the appropriate $\ethreshold$ may vary 
%well 
from one DNN model or dataset to another. 
Also, \sysname should systematically find the appropriate $\ethreshold$, relieving the practitioners of the additional burden of finding an appropriate $\ethreshold$ on their own. 
Thus, \sysname discovers an appropriate $\ethreshold$ by performing a binary search over $\ethreshold$ and finding the highest $\ethreshold$ such that the approximation corresponding to $\ethreshold$ incurs a negligible inference accuracy loss. This is sound due to the monotonic relationship between approximation error and inference accuracy. 
Lastly, one challenge of this approach is checking if the inference accuracy loss is small at each step of the binary search. We describe a solution to this challenge next. 

\subsubsection{Measuring accuracy loss} \label{sec:accuracy-loss-measure}
It is difficult to analytically find the optimal $\ethreshold$ that minimizes inference  accuracy loss while having reduced performance overhead.
%\fixme{One of the crucial components of $\candidatelistgenerationalgo$ is for finding the appropriate $\ethreshold$ %is to tell apart if the generated $\fmpc$ renders a negligible accuracy loss.} 
%Unfortunately, it is difficult to tackle this challenge analytically. 
% before we have the client's input during $\fmpc$ generation process. 
We attempt to handle this challenge empirically by relying on well-known \textit{closed-world} assumption used in machine learning (i.e., for each testing class enough representative examples are available in the training dataset). 
More specifically, we replace the original $\fx$ with the generated MPC-friendly approximation $\fmpc$ and calculate the inference accuracy over the training dataset. 
We call this  inference accuracy 
$\secureInfAcc$ and compare it with the plaintext inference accuracy $\plainTextAcc$ which uses the original $\fx$ over the same training dataset. % for evaluating the accuracy loss of $\fmpc$. 
If $(\plainTextAcc - \secureInfAcc)/\plainTextAcc \le \negligibleaccuracy$, we consider $\fmpc$ to be accurate enough, where $\negligibleaccuracy$ is a small value representing the accuracy loss the practitioner can tolerate for switching to secure inference  from plaintext inference. % for secure inference. 

%
% More specifically, we run secure inference using $\fmpc$ over the training dataset and compare the secure inference accuracy $\eta_1$ with $\eta_0$ --- the inference accuracy using $\fx$ over the same training dataset. If $\eta_1 \sim \eta_0$ then we consider the accuracy loss to be negligible.

\subsubsection{Designing $\fcrude$} \label{sec:designing-fcrude} 
% \rcnote{This section needs a bit more TCL}
We also added another DNN model-specific optional optimization. 
Instead of approximating the original $\fx(x)$, we manually 
introduce a crude MPC-friendly approximation of $\fx(x)$, which we call $\fcrude$.\footnote{For simplicity this is not shown in~\figref{algo:candidateListGen} $\candidatelistgenerationalgo$.} 
Then, we use $\candidatelistgenerationalgo$ procedure to approximate $\left(\fx(x) - \fcrude(x)\right)$. The final approximation of an AF would be $\fcrude(x) + \fmpc(x)$.  Note that $\fcrude$ is designed to be simple and linear, making it easy to use with standard MPC libraries. We find this approach significantly improves the approximation procedure
% will generate a list of MPC-friendly approximations of 
% $\fx(x) - \fmpc(x)$ instead of approximating the original 
% $\fx(x)$. 
% And it was also proposed in prior work~\cite{NFGenFanCCS22}. 

For $\silu$ AF, since $\silu(x) = x \cdot \sigmoid(x)$, we can simply 
borrow the 
structure of the MPC-friendly approximation for\\$\sigmoid(x) \approx  \max(0, \min(x + 0.5, 1))$ from~\cite{mohassel2017secureml}. We tweak it slightly to be more precise and multiply it by $x$ to get $\newfcrude{silu}$ as shown in~\eqnref{eqn:crude-silu}.
         \begin{equation}
         \label{eqn:crude-silu} 
         \newfcrude{silu}(x)  = x \cdot \max\Big(0, \min(6x + 0.5, 1)\Big) \\
    \end{equation}
    % \rcnote{How did you get these numbers?}
    
For $\gelu$ AF, since $\gelu(x) \approx x \cdot \sigmoid(1.702x)$, similarly we can write crude MPC-friendly approximation of $\gelu$ AF by leveraging the same structure of MPC approximation for $\sigmoid$ as shown in~\eqnref{eqn:crude-gelu}.
    \begin{equation}
        \label{eqn:crude-gelu}
        \newfcrude{\gelu}(x)  = x \cdot  \max\Big( 0, \min(10x, 0.5)\Big)
    \end{equation}
    Since $\mish$ cannot be expressed easily in terms of $\sigmoid$, we denote crude MPC 
    friendly approximation of it by $\relu$ as shown in~\eqnref{eqn:crude-mish}.
    \begin{equation}
        \newfcrude{\mish}(x) = \max(0,x)
        \label{eqn:crude-mish}
    \end{equation} 

\subsubsection{Performing interpolation} \label{sec:interpolation}
We  interpolate $f(x)$  between range $[\alpha, \beta]$  by a $\maxdegree$-degree polynomial $\poly$ (\eqnref{eqn:polynomial})
using the  $\interpolate$ procedure. 
To find the best performing $\poly(x)$, similar to $\nfgen$, we adopt 
Chebyshev interpolation~\cite{chebinterpolation}  
over other alternatives, such as cubic spline or uniform polynomial. 
This is due to an established fact in the area of function approximation theory~\cite{miranda1996lecture} that Chebyshev  polynomial interpolation generally has superior performance to 
cubic spline or uniform polynomials interpolation when $f(x)$
involves non-linear operations such as $e^{-x}, \ln, \tanh$, as it is the case with complex AFs shown in~\figref{fig:actfuncs}.

% is smooth and monotonic (c.f.,~\cite{miranda1996lecture} Table 5.1) as it is in our case for complex AF used in DNN models~(\figref{fig:actfuncs}).

\paragraph{$\candidatelistgenerationalgo$ procedure.}
Now we can piece together the above-mentioned techniques and describe the procedure to approximate 
$\fx$ within region $[s, e]$ using a number of piece-wise polynomials 
in detail (as shown in  $\candidatelistgenerationalgo$~\figref{algo:candidateListGen}).
First, we set a step size $\stepsize$,  $\alpha \gets s$, and $\beta \gets \alpha + \stepsize$. 
Then at each step, we increase the pointer $\beta$ by $\stepsize$.   
Before increasing $\beta$, we check if 
the adjusted approximation error $\ethreshold'$ in the region $[\alpha, \beta]$ is more than the  expected approximation error $\ethreshold/\npieces$.  

If this is the case, we approximate the region $[\alpha, \beta]$ using a polynomial $\poly$ using the Chebyshev interpolation algorithm, add that polynomial piece to $\fmpc \gets \fmpc \bigcup \{\poly\}$,  and update $\alpha$ to $\beta$.
Next, we update $\beta$ by $\stepsize$, and again perform the above-mentioned check until we have approximated 
the whole region $[s, e]$.

% \vspace{-0.8cm}
\subsection{Finding  Efficient Approximation}
\label{sec:time}

Now that we can generate MPC-friendly approximations $\fmpc$ 
using $\candidatelistgenerationalgo$ procedure that have negligible accuracy loss for a given $\langle \npieces, \maxdegree, \ring \rangle$, one can search over all possible values of  $\langle \npieces, \maxdegree, \ring \rangle$ and select the  $\fmpc$ that is computationally more efficient. We 
% abuse the notation slightly and 
use $\solution$ to represent $\langle \npieces, \maxdegree, \ring \rangle$. \changed{We also use $\Time(\fmpc')$
to represent the average time it takes to complete secure inference with the approximation  $\fmpc'$ generated % by 
using $\theta'$ in the approximation process.
The accuracy loss can be presented by $\AccuracyLoss(\fmpc', \fx) = (\plainTextAcc-\secureInfAcc)/\plainTextAcc$; where
$\plainTextAcc$ is the accuracy when we use the complex AF  $\fx$ as it is (i.e., plaintext accuracy), and 
$\secureInfAcc$ is the accuracy when
we replace the complex AF with its MPC-friendly approximation $\fmpc'$ (i.e., secure inference accuracy). 
%Thus, $\negligibleaccuracy = (\plainTextAcc-\secureInfAcc)/\plainTextAcc$ gives the inference accuracy loss introduced by MPC-friendly approximations.
}

% , and the accuracy loss }

Unfortunately, because of performing the binary search to find the appropriate $\ethreshold$, $\candidatelistgenerationalgo$ becomes time-consuming. 
This is because determining if the accuracy loss is negligible at each step of binary search with reasonable confidence requires performing inference over the large training dataset (as explained in~\secref{sec:accuracy-loss-measure}), and it makes exhaustively iterating over all possible values of $\solution$ from the search space $\Theta$ infeasible. 

In this work, we treat this problem of finding optimal $\solution$ used for the approximation to 
balance performance overhead and accuracy loss, as a constraint optimization problem (COP).
Roughly, this means, we find a $\solution' \in \Theta$ that minimizes the average inference time 
under the constraint that the accuracy loss is less than a specified threshold $\negligibleaccuracy$.

Concretely, for a given $\fx$, we want to solve the following optimization problem. 
%~\eqnref{eqn:optimization}.
\bne 
\theta \gets \argmin_{\theta' \in \Theta} \Time(\fmpc') \text{ s.t }\AccuracyLoss(\fmpc', \fx) \le \negligibleaccuracy
\label{eqn:optimization}
\ene
Here $\negligibleaccuracy$ is the specified maximum accuracy loss threshold we can tolerate, and $\fmpc'$ is the  MPC-friendly approximation of $\fx$ generated by \sysname  using $\theta'$.
\changed{
Now to solve this COP problem in~\eqnref{eqn:optimization}, we devise 
% an application-specific 
a search technique based on simulated annealing (SA)~\cite{johnson1989optimization}. 
SA is a general framework to tackle COP. Briefly, SA starts with an initial solution, generates new neighboring solutions relative to the current solution, and probabilistically decides between moving to the new solution or staying with the current solution for fixed number of iterations.} 
One advantage
of sketching SA-based searching for optimal solution $\solution$ is that SA is gradient-free --- suiting our needs --- overcoming the difficulty to underpin an analytical formula of $\nabla_{\solution = \langle \npieces, \maxdegree, \ring \rangle} \candidatelistgenerationalgo(\cdot)$.  
That being said, other gradient-free searching techniques may also work as well~\cite{eren2017introduction}, and we detail some additional discussions about this in~\appref{app:minmax}. 

One important characteristic of SA---we need to model for this case---is how to avoid being trapped in a local suboptimal solution.
To this extent, we follow suggestions from prior work~\cite{deb2012optimization,hwang1988simulated}, and \emph{probabilistically} move towards a new solution $\solution_{\step}$ even if $\solution_{\step}$ is computationally less efficient approximation than the current best solution ($\tcur$).
%  as shown in~\eqnref{eqn:update-cur}.

More precisely, if  at $\step$-th iteration, we denote the MPC-friendly approximation from $\solution_{\step}$ as $\fmpc^{\step}$, then we always update our current best solution $\tcur$ to 
$\solution_{\step}$ if $\fmpc^{\step}$ is computationally more efficient than $\fcur$ (i.e., $ \Time(\fcur) > \Time(\fmpc^{\step})$). Otherwise, we update $\tcur$ to $\solution_{\step}$ with a certain acceptance probability.
This probability depends on two factors. First, the temperature 
at $\step$-th iteration called $T_\step$ --- which is initially high, meaning we have a high tendency to accept a solution computationally less efficient, but after a few more iterations $T_\step$ decreases and so does our tendency to accept a computationally less efficient solution.
Second, the amount of computation less efficient $\fmpc^{\step}$ is compared to $\fcur$. 
In other words,  
we accept $\solution_{\step}$ when $ \exp(\psi/T_\step) > \randomNumber$ is true. Here $\psi$ represents the computational efficiency of  $\fmpc^\step$ over the current approximation $\fcur$  expressed as
$\psi \gets \Time(\fcur)-\Time(\fmpc^\step)$, and $r$ is a randomly chosen number given by $r \getsr U_{[0, 1]}$. 

% % \bnm 
% \begin{equation}
% % \tabfontsize
% \tcur = \begin{cases}
% \solution_{\step} & \textsf{if } \Time(\fcur) > \Time(\fmpc^{\step}) \text{ or } \\
% & \exp\bigg(\big(\Time(\fcur)-\Time(\fmpc^{\step})\big)/T_\step\bigg) > \randomNumber \\ 
% & \mytab \textsf{where } r \getsr U_{[0, 1]}\\ 
% \tcur & \textsf{otherwise} 
% \end{cases}
% \label{eqn:update-cur}
% % \enm 
% \end{equation}

% Here 
% $\Time(\cdot)$ can be logged simultaneously when we run the secure inference using $\fmpc$ inside $\candidatelistgenerationalgo$. 

We have to design two more parameters carefully. One is the neighborhood generation heuristic for $\solution$, and the other is setting a cooling schedule for the temperature $T_\step$. 
Without careful handling of these two parameters
SA may lead to undesired approximations~\cite{ben2004computing}. 

\paragraph{Neighbour generation heuristic.}
At iteration $\step$, 
we generate a new neighbor  $\solution_\step = \langle \npieces', \maxdegree', \ring' \rangle$ from  $\solution = \langle \npieces, \maxdegree, \ring \rangle$
in the following way: for $\npieces', \maxdegree'$ we randomly sample two integer numbers $z_1, z_2\in\Z$ 
from a probability distribution having density function $p(x=z)= (1/3)\cdot2^{-|z|}$
such that %$P: \mathbb{Z} \rightarrow [0, 1]$ with probability density function defined as 
% $P(X = z) = 1/3\cdot2^{|z|}$ 
and set $\npieces' \gets \npieces + z_1$ and  $\maxdegree' \gets \maxdegree + z_2$. % We also decide the sign for $z_1$ and $z_2$ based on a fair coin toss. 
This means that the chances of moving further away from the current value  $\npieces$ and  $\maxdegree$ decreases exponentially.
%while the chances of selecting values close to current $\npieces, \maxdegree$ is higher. 
% \fixme{Probabbly need to divide this by Riemann zeta function(1) to make probabilities sum up to 1?}. 

% Handling $\ring_{\resolution,\decimal}$ requires  more consideration. 
To specify a $\ring$, we need two numbers: \emph{i)} the size of the ring used in MPC library (denoted by $\resolution$), and \emph{ii)} the number of last bits to represent the fractional parts (denoted by $\decimal$).
Typically, MPC libraries use $\ring$ sizes of $\{128, 84, 64, 32\}$. %~\cite{peter-paper}. 
We randomly sample a ring size from these for $\resolution$, and regarding values of $\decimal$ we set it to $\decimal \gets \floor{\resolution/\gamma_2}$ where $\gamma_2$ is randomly sample from  $\gamma_2 \in_{R}
%\{3/2, 2, 5/2,3, 7/2, 4\}
\{1.5, 2, 2.5, 3, 3.5, 4\}$.
% \minote{$z_1$ and $z_2$ should be both positive and negative.}
% cite the paper Peter shared on mixed protocols

\medskip\noindent\textbf{Cooling schedule.}
As for the cooling schedule, we adopt the classical logarithmic series  $T_\step \gets \chi_0/\log(\step + 1)$ 
at $\step$-th iteration
following Hajek et al.~\cite{hajek1988cooling}. This choice ensures that initially, $T_\step$ would be high, thereby increasing the chances of accepting a computationally less efficient approximation during the early iterations. But as the number of iterations increases, $T_\step$ progressively decreases, lowering this chance. 
% (i.e., lowering the chances of accepting a computationally less efficient approximation).
We simply set $\chi_0 = 0.2$ for all of our experiments, yielding $T_{1} \approx 0.67$ and  $T_{10} \approx 0.2$. 

We show the pseudocode for finding computationally efficient approximation  $\ApproximationAlgo$ and the procedure for generating neighbors at each iteration $\generateNeighbour$ in~\figref{algo:candidateListGen}.
% }
% \vspace{-0.4cm}
\section{Experimental Evaluation} \label{sec:experiments}
We conduct experiments to address the following questions:
\begin{newenum}
    \item{\textit{Model Accuracy}~(\secref{sec:accuracy}):}
    What is the impact on model inference accuracy of using MPC-friendly activation functions $\fmpc(x)$ generated using our scheme $\sysname$ and 
    other existing approaches~\cite{NFGenFanCCS22,liu2017oblivious,li2022mpcformer,rathee2021sirnn}?
    \item{\textit{Inference Time}}
(\secref{sec:performance}):
    What is the inference time overhead of $\sysname$ compared to $\nfgen$~\cite{NFGenFanCCS22}  as the number of hidden layers increases? 
    %without losing any significant loss in inference accuracy?
\end{newenum}

% Implementation and experimental setup details are described in~\secref{sec:implementation-details} and ~\secref{sec:image-classification}, respectively.

\subsection{Implementation Details} \label{sec:implementation-details}

\paragraph{Our Scheme.}
We implement our scheme 
using \funcfont{Python 3.8} in about $800$ lines of code. We approximate the region between $x \in [-5, 5]$ for all activation functions (AFs) as beyond that region, they can be easily approximated using simple polynomials. Also, we use  \texttt{SymPy}~\cite{sympy} %\footnote{\url{https://docs.sympy.org/latest/index.html}} 
library for the majority of mathematical operations, including calculating the approximation error between a given region of a polynomial using~\eqnref{eqn:mean_approximation_error} and performing  Chebyshev interpolation as mentioned in~\secref{sec:interpolation}. Our source code is
publicly available~\cite{compact-github}.

% \begin{table*}[t]
%     \centering
%         \begin{tabular}{@{}cll|rrr|rrr|rrr|rrr@{}}
%         \toprule 
%         \multirow{2}{*}{\# of parties} & \multirow{2}{*}{Protocol} &  \multirow{2}{*}{Scheme} & \multicolumn{3}{c|}{$\taskA$} & \multicolumn{3}{c|}{$\taskB$} & \multicolumn{3}{c|}{$\taskC$} & \multicolumn{3}{c}{$\taskD$} \\  \cmidrule{4-15}
%         & & &  $\silu$&$\gelu$&$\mish$&$\silu$&$\gelu$&$\mish$& $\silu$&$\gelu$&$\mish$&$\silu$&$\gelu$&$\mish$ \\

%         \midrule
%         \bottomrule
%         \end{tabular}
%         \caption{Comparison of inference time (ms)
%         of three activation functions over four different tasks  (referred in~\figref{tab:classification-accuracy}). 
%         }
%         % specify the field size. 
%         \label{tab:performance}
%         \end{table*}

Our scheme requires testing if the generated approximation has negligible accuracy loss by checking $(\plainTextAcc - \secureInfAcc)/\plainTextAcc \le \negligibleaccuracy$ (as described in~\secref{sec:accuracy-loss-measure}).
% We set $\negligibleaccuracy = 10^{-2}$ for our experiments.
We also configure $\ApproximationAlgo$ with ten iterations ($\step_{max} = 10$) to find a computationally efficient approximation and set $\chi_0=0.2$. 
For the initial solution $\theta_0 = \langle \npieces_0, \maxdegree_0, \ring_0 \rangle$, we set  $\npieces_0 = 10^4$ and  $\maxdegree_0 = 10$ (default parameters taken from $\nfgen$~\cite{NFGenFanCCS22}). 
For $\ring_{\langle \resolution_0, \decimal_0 \rangle}$, we use 
$\langle \resolution_0, \decimal_0 \rangle = \langle 128, 64 \rangle$ --- a popular choice of ring size by many MPC libraries.
With this configuration, $\ApproximationAlgo$ took less than 25 minutes on commodity hardware to finish the four tasks and three complex AFs we detail in~\secref{sec:image-classification}. 
% Multiprocessing can also be used to further accelerate this time.  
\tabref{tbl:configurations} shows the appropriate $\npieces, \maxdegree, \ring$ we find via $\ApproximationAlgo$ for all tasks and complex AFs.

% We note \sysname can also be implemented by 
% adding our DNN specific optimizations (as discussed in~\secref{sec:our-piece-wise-poly}) on top of $\nfgen$ code base, and we expect to see similar efficacy over  $\nfgen$ --- to what Fan et al. propose in~\cite{NFGenFanCCS22}. 
% However, when $\nfgen$ was publicly released~\cite{nfgen-code}, we were already towards the end of the development cycle of our code base, and hence, we did not reuse their code base for 
% % implementing techniques employed by 
% \sysname.

\paragraph{Other Approaches.}
We consider four state-of-the-art approaches for comparison:
$\nfgen$~\cite{NFGenFanCCS22}, MiniONN~\cite{liu2017oblivious}, MPCFormer~\cite{li2022mpcformer} and SIRNN~\cite{rathee2021sirnn}. 
Additionally, we consider a rudimentary base approach: replacing the complex AF with a popular MPC-friendly AF $\relu$. 
We consider this approach as $\relu$ is relatively MPC-friendly because it can be computed using only two piece-wise polynomials.

For $\nfgen$, we add a wrapper class to the author's open-source implementation to measure the inference accuracy and computational overhead for the four tasks. Besides that, we keep their implementation unchanged --- using  $\MaxApproxError$ (\eqnref{eqn:max_approximation_error}) to measure the approximation error, setting $\ethreshold = 10^{-3}$, $\maxdegree = 10$, and $\npieces = 10^4$.
% n, and f?
% https://github.com/Fannxy/NFGen/blob/main/src/NFGen/main.py
Liu et al.~\cite{liu2017oblivious} describe an approach called MiniONN for generating MPC-friendly approximations of $\sigmoid$ 
AF. Since there is no publicly available implementation of MiniONN, we implement it ourselves 
based on the description given in~\cite{liu2017oblivious}
%to the best of our abilities 
and extend the approach to generate MPC-friendly versions of complex AF $\fx \in \{\silu, \mish, \gelu\}$. Further details of MiniONN, and our extension are discussed in~\appref{app:MiniONN}.

MPCFormer~\cite{li2022mpcformer} approximates $\gelu$ 
using an MPC-friendly polynomial given by $\gelu(x)= 0.125x^2 + 0.25x + 0.5$. This approximation was motivated by the need to perform secure inference for transformer-based DNN models where $\gelu$ activation is used extensively.
Since Li et al. ~\cite{li2022mpcformer} did not provide any recipes that could be generalized directly to other AF, we only compare the accuracy and computational overhead for $\gelu$.

Lastly, Rathee et al.~\cite{rathee2021sirnn} present a library called SIRNN that computes complex mathematical operations (e.g., $e^{x}, \ln(x), \frac{1}{x}$) securely using a combination of 
lookup tables and numerical methods (e.g., Goldschmidt's iterations). 
Thus, complex AFs can be computed sequentially by performing 
the aforementioned operations and combining the intermediate results using $\addmathoperation, \mulmathoperation, \comparemathoperation$ operators to evaluate $\fx$. Recently, Hao et al.~\cite{hao2022iron} extended their approach to computing $\gelu$ AF efficiently by reducing one network call. Nevertheless, this work uses the open source implementation of SIRNN~\cite{sirnn-lib}.

% \definecolor{electriclime}{gray!40}

\begin{table*}[t]
    \centering
    \tabfontsize
    \resizebox{\linewidth}{!}
{
    \begin{tabular}{llllr@{\hskip 10mm}rrrrrr}
    \toprule   
    % \midrule 
    \multirow{2}{*}{Task Name} & \multirow{2}{*}{Model$^\dagger$} & \multirow{2}{*}{Dataset} &  \multirow{2}{*}{$\fx$} &
    \multirow{1}{*}{\% Plaintext}  & \multicolumn{6}{c}{\% Accuracy loss$^\S$} \\     
    & &  &   & \multirow{1}{*}{accuracy} & \multirow{1}{*}{$\relu$} & NFGen  & MiniONN & MPCFormer & SIRNN & $\sysname$\\ 
    \midrule  
    \multirow{3}{*}{$\taskA$} & \multirow{3}{*}{FCN} & \multirow{3}{*}{MNIST} & $ \silu$ 
    & 98.73 &   2.31	& \cellcolor{electriclime} 0.43 & 20.88	& \emph{n/a} &	2.37	& \cellcolor{electriclime} 0.17 \\  
     & & & $\gelu$ & 
     98.45 &   1.54 & \cellcolor{electriclime}	0.23 &	42.31	& \cellcolor{electriclime} 0.18	& 1.32	& \cellcolor{electriclime} 0.97 \\
     & & & $\mish$ & 99.07 & 2.68 & 	\cellcolor{electriclime} 0.19	& 30.41	& \emph{n/a} 	& \cellcolor{electriclime} 0.95 &	\cellcolor{electriclime} 0.06 \\
     \midrule

    \multirow{3}{*}{$\taskB$} & \multirow{3}{*}{ConvNet} & \multirow{3}{*}{CIFAR-10} & $\silu$ &  
    86.53 & 49.80 & 	\cellcolor{electriclime} 0.51 & 	18.50	& \emph{n/a}	& 2.58	& \cellcolor{electriclime} 0.49\\ 
    & &  & $\gelu$ &  87.11 & 45.66	& \cellcolor{electriclime} 0.64	& 30.04	& 7.07	& 4.01 & 	\cellcolor{electriclime} 0.25 \\
    & &  & $\mish$ &  89.30 & 57.07 & \cellcolor{electriclime}	0.27 & 	57.07 & 	\emph{n/a}	& 13.64	& \cellcolor{electriclime} 0.11  \\
    \midrule 

    \multirow{3}{*}{$\taskC$} &  \multirow{3}{*}{ResNet9} & \multirow{3}{*}{ImageNet-1K} & $\silu$ &
    72.89 & 98.39	& 1.36 & 	27.12 & 	\emph{n/a} & 	10.59 & 	\cellcolor{electriclime} 0.91 \\ 
    & &  & $\gelu$ & 75.43 & 77.66	& \cellcolor{electriclime} 0.05	& 36.21	& 9.43	& 6.68	& \cellcolor{electriclime} 0.03 \\
    & &  & $\mish$ & 75.78 & 98.97	& \cellcolor{electriclime} 0.61 & 	39.89 & 	\emph{n/a}	& 16.31	& \cellcolor{electriclime} 0.55\\
     \midrule
    
    \multirow{3}{*}{$\taskD$} & \multirow{3}{*}{EfficientNetB0} & \multirow{3}{*}{CelebA-Spoof} & $\silu$ &  
    90.87 & 71.72 & 	\cellcolor{electriclime} 0.14 & 	4.27 & 	\emph{n/a}	& 1.75 & 	\cellcolor{electriclime} 0.08 \\ 
    & & & $\gelu$ & 92.19 & 75.94 & \cellcolor{electriclime}	0.20 & 	9.75 & \cellcolor{electriclime} 0.09	 & \cellcolor{electriclime}	0.48 &  \cellcolor{electriclime} 0.77	 \\
    & & & $\mish$ & 92.23 & 77.71	& \cellcolor{electriclime} 0.53 & 	1.32 & 	\emph{n/a} & 	1.78 & \cellcolor{electriclime}	0.66 \\
    % \midrule
    \bottomrule  
    \end{tabular}
}
\vspace{0.5em}
    \begin{flushleft}
        % \tabfontsize
        \emph{n/a} =  MPCFormer does not propose MPC friendly approximation for  $\silu, \mish$. \\ 
        % $^\dag$We set the approximation error threshold ($\ethreshold$) to $10^{-1}$ for $\sysname$ and $10^{-3}$ for NFGen. Our experimentation with different values of $\ethreshold$ and their effect on inference accuracy is shown in~\secref{sec:performance-accuracy-tradeoff} \\ 
        \textsuperscript{\S}Accuracy loss is reported by comparing the
        inference accuracy $\plainTextAcc$ and $\secureInfAcc$ obtained using AF $\fx$ and $\fmpc$, respectively.
        %and accuracy $(\eta_2)$ using the approximated AF $(\fmpc)$, that is
        Accuracy loss $= (\plainTextAcc - \secureInfAcc)/\plainTextAcc$, and reported in percentage ($\%$). Accuracy losses of $< 10^{-2}$ or $< 1\%$ are highlighted in gray. \\ 
        $^\dagger$For all models, batch normalization is used before each activation layer. 
    \end{flushleft}
    \caption{Inference accuracy of MPC-friendly approximation of three complex activation functions (AF) for four different tasks using state-of-the-art approaches. Except for $\nfgen$ and our approach other DNN-specific approaches show a significant drop in inference accuracy if we use their generated MPC-friendly version of complex AF. We compare the performance overhead of our approach with $\nfgen$ in~\secref{sec:performance} and show results in~\tabref{tab:performance}.
    }
    \label{tab:classification-accuracy}
   % \vspace{-3em}
   % \vspace{-0.5cm}
\end{table*}

\vspace{0.2cm}
\subsection{Experimental Setup} \label{sec:image-classification}
% \paragraph{Task details.} 
To demonstrate that inference accuracy and performance overhead 
is negligible
for secure inference using 
our scheme, %$\sysname$, 
we  consider four state-of-the-art image classification 
tasks as shown in~\tabref{tab:classification-accuracy} and three complex activation functions (AF) $\fx \in \{\silu, \gelu, \mish\}$.
We train the four models corresponding to each complex AF for each task.  
While training these models, we preserve
the widely use parameters as proposed in the 
literature for all models (e.g., the overall architecture of the model, 
\# of epochs, learning rate, optimizer, etc.) --- including a batch normalization layer before inputs are being fed to  
complex activation functions of each hidden layer as illustrated in~\figref{fig:normalization}. 

Below, we provide brief details about these four classification tasks, and further details are in~\appref{app:dnn}.

\paragraph{Four classification tasks.}
For the first task, 
we consider a simple classification task of MNIST dataset~\cite{lecun1998gradient} using 
a three-layer deep fully connected network (FCN) with one input, output, and hidden layer. MNIST dataset contains  70 thousand $28 \times 28$ handwritten digits grey images, and the three-layer deep FCN achieves
close to $0.99$ training accuracy for the three complex AFs. We refer to this task as 
\taskA in the paper.
% More details about MNIST and its training accuracy.
Next,  we move towards a more complex classification task of 
CIFAR-10 dataset~\cite{krizhevsky2009learning} --- which we refer to as \taskB.

CIFAR-10 consists of 
60 thousand $32\times32$ color images with six thousand images per 10 classes.  
For performing classification on this dataset, we use a
a convolutional  
neural network~(ConvNet)~\cite{o2015introduction} with five hidden layers
and train it over the 50 thousand training images of CIFAR-10 dataset using three different complex AFs.
For the third task, we consider performing classification on ImageNet-1K dataset which has been one of the challenging benchmark datasets in image classification ~\cite{deng2009imagenet}.   We refer to this task as \taskC in this paper.

The ImageNet-1K  dataset contains around one million annotated images with 50 thousand validation images and 100 thousand test images. We use a deep 
residual neural network (ResNet9)~\cite{he2016deep} model having eight hidden layers over the training images for 50 epochs for three complex activation functions and achieved a validation accuracy of around 0.74. 
Lastly, we perform experiments to detect spoofed images in CelebA-Spoof~\cite{zhang2020celeba} dataset.
We refer to this task as \taskD in this paper. 

CelebA-Spoof is a large-scale face anti-spoofing dataset used to train anti-spoofing DNN models. CelebA-Spoof contains 625 thousand facial images from around 10 thousand subjects with each image having 43 attributes; 40 of them correspond to indicating facial components of real images and three of them correspond to attributes of spoofed facial images.
For training, we perform an 80\%-20\% split of the CelebA-Spoof dataset and  
adopted the EfficientNetB0~\cite{tan2019efficientnet} model, which is the state-of-the-art top-performing anti-spoofing detection model and winner of the CVPR challenge of detecting spoofed face images~\cite{liu20213d}. 
The EfficientNetB0 model consists of 17 hidden layers, and after training the model for 25 epochs, it achieved a training accuracy of 0.98.

\paragraph{Machine specification.}
We train the models on a Linux machine with an Intel Core i9 processor having 128~GB RAM and Nvidia GTX 1080 GPU. The training split of each dataset is used for training the models. After the training is completed, we save these models.  We assume $\modelowner$ holds these saved models and the weights of these models 
are $\modelinput$ which $\modelowner$ 
does not want to reveal to $\client$ while performing secure inference.  
We simulate the $\client$'s input $\clientinput$ 
using the testing split of the corresponding datasets for each task.

%\vspace{-1cm}
\subsection{Model Accuracy} \label{sec:accuracy} 
% \vspace{-0.75cm}
We first measure the inference accuracy of the trained models over the testing split of the dataset by using the (non-MPC-friendly) complex activation functions as it is and refer to it as \textit{plaintext accuracy} $(\plainTextAcc)$. 
Then, we replace the complex activation function with its MPC-friendly approximation generated by different approaches and measure its inference accuracy $(\secureInfAcc)$. Thus, $\negligibleaccuracy = (\plainTextAcc-\secureInfAcc)/\plainTextAcc$ gives the inference accuracy loss introduced by MPC-friendly approximations. 
% compared to the plaintext one.
\tabref{tab:classification-accuracy} 
shows the inference accuracy loss (in percentage) for \sysname generated MPC-friendly approximation 
and other state-of-the-art approaches generated MPC-friendly approximation~\cite{NFGenFanCCS22,hao2022iron,liu2017oblivious} --- for each task across the three complex AFs $\silu$, $\gelu$, $\mish$.

Now we discuss the inference accuracy loss for different approaches, and throughout the discussion, we conservatively consider accuracy loss negligible if $\negligibleaccuracy < 10^{-2}$.

\paragraph{$\relu$ based approach.}
% For a rudimentary approach, we consider replacing the complex activation function with $\relu$ during inference as $\relu$. Primarily we consider this approach as $\relu$ is already MPC-friendly because it can be computed using only two piecewise polynomials. 
We observe that although for the first \taskA task, the inference accuracy loss is within 1.54\%-2.68\% for the last three tasks 
% \taskB, \taskC, \taskD 
accuracy loss is higher --- at least 45.66\% --- making this approach unsatisfactory. 

\paragraph{SIRNN~\cite{rathee2021sirnn}.}
For SIRNN, we observe that for \taskA task, we observe less significant accuracy loss ($0.95\%$--$2.37\%)$.
Furthermore for \taskD the accuracy does not degrade too much --- by $0.48\%$--$1.78\%$. 
However, for \taskB and \taskC task the accuracy degradation is higher --- suffering from an accuracy loss 
of $2.58\%$--$16.31\%$. 

We  hypothesize such accuracy degradation is primarily 
due to two reasons:  
1) intermediate steps overflowing in the fixed point representation, 
and 2) errors introduced in one layer propagating to subsequent layers and 
accumulating in the process. %while using that result as input for another complex math operation. 
This further motivates the need to  take a piece-wise polynomial approximation-based 
approach for designing MPC-friendly approximation of complex AF when state-of-the-art DNN models are used, confirming findings from prior work~\cite{NFGenFanCCS22}.

\paragraph{MPCFormer~\cite{li2022mpcformer}}
% MPCFormer --> optimized for transformers architecture
For MPCFormer, we observe a negligible accuracy loss for \taskA and \taskD tasks of 0.18\% and 0.09\% respectively. However, similar to SIRNN, it exhibits a non-negligible 
accuracy loss of 9.4\% and 7.07\% for \taskB and \taskC task respectively. 
We suspect this is because $\gelu$ activation approximation by MPCFormer 
relies on \emph{knowledge distillation} (KD)~\cite{hinton2006fast} -- which is essentially fine-tuning the sequence-to-sequence-based pre-trained model for efficiency. In absence of KD, a simple plug-and-play replacement of polynomial approximation of $\gelu$ activation proposed by MPCFormer does not work well.   

\paragraph{MiniONN~\cite{liu2017oblivious}.}
For MiniONN, we observe that the inference accuracy loss becomes significant when we use their recipe to generate a friendly approximation of complex AFs $\silu$, $\gelu$, $\mish$.
The accuracy loss becomes catastrophically high, especially for \taskC, ($27.12\%$--$39.89\%$). 
This shows that although the recipe proposed by MiniONN does not show accuracy degradation
for $\sigmoid$ AF for simplistic logistic regression models, there is a generalization gap when such recipes are used 
for DNN models trained on diverse datasets involving complex AFs.  

\paragraph{$\sysname$ and $\nfgen$~\cite{NFGenFanCCS22}}
We observe that for all tasks, in general, \sysname and \nfgen generated MPC-friendly approximations have 
negligible accuracy loss of $< 1\%$. For one instance, though,  \taskC task involving $\silu$ AF, $\nfgen$ has an accuracy loss of 1.36\% --- marginally higher than the aforementioned threshold. 
When comparing the two approaches, generally, $\sysname$ generated approximation has lower accuracy loss, except for two instances showing a slight deviation. 
The first one is \taskA task involving $\gelu$ (0.37\% vs 0.23\%) and \taskD task involving $\mish$ AF (0.66\% vs 0.53\%). 

\paragraph{Results summary.} We conclude from these experiments that $\nfgen$ and $\sysname$ are resistant to significant accuracy loss --- when we use their generated MPC-friendly approximation instead of the original complex AF --- compared to other approaches we consider. Keeping that in mind, we can now investigate the next important aspect of secure inference, measuring performance overhead. We narrow down our experiments to $\nfgen$ and $\sysname$ --- excluding the other approaches --- as their accuracy loss is significantly high. 
  
% \vspace{-0.6cm}
\subsection{Inference Time} \label{sec:performance}
% \fixme{machine specification for inference time.}
We benchmark the inference time of  $\nfgen$ and $\sysname$ to measure the performance overhead. 
While benchmarking, we instantiate each party in the protocol by machines running on commodity-type hardware --- having an Intel Core i7 processor with 64~GB RAM and connected over a 252 Mbits/sec network link.
We use the average inference time for a single image calculated over the testing split of the datasets and include both computational and communication costs while reporting the results. 
% We use the testing split of the dataset while reporting the inference time.

We consider two state-of-the-art MPC libraries~\cite{aby3peter, rathee2020cryptflow2} designed for secure inference --- one for a 2PC scenario and the other for a 3PC scenario (as described earlier in~\secref{sec:problem-formulation}). 

\begin{table}[t]
    \tabfontsize
    \resizebox{\linewidth}{!}
    {
    \centering 
    \begin{tabular}{@{}lrlrrr@{}}
        \toprule
        Task Name  & \# HLs$^\dag$ & $\fx$ & NFGen & Ours & \textbf{Speedup} \\ \midrule 
         \multirow{3}{*}{\taskA} & \multirow{3}{*}{1} & $\silu$ 
        & 40 & 43 & 0.93$\times$ \\ 
        & & $\gelu$ & 35 & 32 & 1.09$\times$ \\
        & & $\mish$ & 52 & 49 & 1.06$\times$ \\ \midrule 
        \multirow{3}{*}{\taskB} & \multirow{3}{*}{5} & $\silu$ 
        & 114 & 58 & 1.96$\times$ \\ 
        & & $\gelu$ & 194 & 94 & 2.05$\times$ \\
        & & $\mish$ & 117 & 62 & 1.89$\times$ \\ \midrule 
        \multirow{3}{*}{\taskC} & \multirow{3}{*}{8} & $\silu$ 
        &  359 & 102 & 3.52$\times$ \\ 
        & & $\gelu$ & 446 & 106 & 4.17$\times$ \\
        & & $\mish$ & 473 & 104 & 4.52$\times$ \\ \midrule 
        \multirow{3}{*}{\taskD} & \multirow{3}{*}{17} & $\silu$ 
        & 204 & 47 & 4.34$\times$ \\ 
        & & $\gelu$ & 221 & 45 & 4.91$\times$ \\
        & & $\mish$ & 195 & 41 & 4.75$\times$ \\ 
        \bottomrule
\end{tabular}
    }
    \begin{flushleft}
    % \vspace{0.01cm}
    $^\dag$ \# HLs =  Number of hidden layers. 
    \end{flushleft}
    \caption{
        Comparison of inference time (ms) of three activation functions ($\fx$) over four different classification tasks for $n=3$ computing servers using \abythree library. 
        %DNN model used in $\taskA{}$ task has only one hidden layer (\# HLs=1), the performance of $\nfgen$  is similar to \sysname. However, for  DNN models with many hidden layers used in the three other tasks, \sysname outperforms $\nfgen$ --- exhibiting a 2$\times$--5$\times$ speedup.  
    }
    \label{tab:performance}
    \vspace{-0.8cm}
\end{table}
\paragraph{3PC results.}
First, for the 3PC scenario, we consider \abythree{}~\cite{aby3peter} that uses replicated secret sharing (SS) based secure inference protocol.
Table~\ref{tab:performance} compares the performance overhead 
of \sysname and $\nfgen$ for the 3PC setting using \abythree{} library.
We observe that $\sysname$ outperforms  $\nfgen$ 2$\times$-5$\times$ for the last three classification tasks involving a high number of layers. However, $\sysname$'s performance efficacy for the first task \taskA is comparable to $\nfgen$---exhibiting similar inference time.
We hypothesize this is because the number of hidden layers is only one for the DNN model used in this 
first task. In contrast, the number of hidden layers for the other classification tasks is 5, 8, and 17 respectively. Because of this, there is a higher chance of the approximation errors introduced in one hidden layer propagating to the next hidden layers. 
Our DNN-specific techniques discussed in~\secref{sec:our-piece-wise-poly} can effectively curb out this approximation error from propagating to the hidden next layers without sacrificing much performance overhead compared to $\nfgen$. Thus,  $\sysname$'s  superior performance   becomes more pronounced as
number of hidden layers becomes high.

\paragraph{2PC results.}
For the 2PC scenario, we consider  CryptFlow2~\cite{rathee2020cryptflow2} another state-of-the-art library for secure inference based on a novel protocol for \emph{millionaries' problem}~\cite{yao1986generate} and division over fixed-point arithmetic.
We experiment with the oblivious transfer (OT) based construction of CryptoFlow2 but believe performance results would also be similar for homomorphic encryption (HE) based construction.
We observe a performance efficiency trend of \sysname with \nfgen similar to the 3PC scenario. We present detailed results of those experiments in~\appref{app:crytpflow2}.

\newcommand{\stackedtimeseries}[8]{
  \begin{tikzpicture}[scale=0.47]
    % \node[] at (3.5, 6.1) {\small #5};
    \begin{axis}[
      cycle list shift=10,
      scaled x ticks=false,
      legend cell align={left},
      legend style = {
        at={(0.02,0.98)},
        anchor=north west 
      },
      smooth,
      axis line style={rounded corners},
      every axis plot/.append style={thick},
      ymajorgrids=true,
      xlabel={\Huge \# of hidden layers (\#HLs)},
      ylabel={\Huge #5},
      xtick pos=left,
      ytick pos=left,
    %   ymin=0,
    %   ymax=85, % when xmax is 100, the graph is a right angle
      ]       
      \addplot table[x={#1}, y = {#2}] {\data};
      \addlegendentry{\LARGE \nfgen ($\ethreshold = 10^{-3}$)};
      \addplot table[x={#1}, y = {#4}] {\data};
      \addlegendentry{ \LARGE \nfgen ($\ethreshold = 10^{-1}$)};
      \addplot table[x={#1}, y = {#3}] {\data};
      \addlegendentry{\LARGE \sysname (Ours)};
      \draw[dashed] (axis cs:10,#6) -- node[left, yshift=-3ex]{\LARGE #8$\times$} (axis cs:10,#7);

    \end{axis} 
  \end{tikzpicture}
}

% mention in our full version of the paper we will experiment with more values of $\ethreshold$.
\begin{figure}[t]
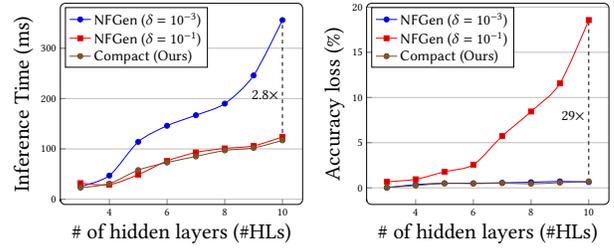

    \centering %\figfontsize
    \pgfplotstableread[col sep = comma]{figures/data/ablation_study.csv}\data
    \stackedtimeseries{layers}{InferencetimeNFGenOriginal}{InferencetimeCompact}{InferencetimeNFGenCrude}{Inference Time (ms)}{117}{356}{2.8}
    \stackedtimeseries{layers}{AccuracyNFGenOriginal}{AccuracyCompact}{AccuracyNFGenCrude}{Accuracy loss (\%)}{0.64}{18.567}{29}
    % \begin{flushleft}
    %     \tabfontsize
    %     $^\dag$ Lower is better for both inference time and accuracy loss.  \\ 
    %     We experiment with two approximation error threshold for \nfgen: i) $\ethreshold =  10^{-1}$ (a crude one we set) and ii) $\ethreshold = 10^{-3}$ (used by Fan et al.~\cite{NFGenFanCCS22}).
    % \end{flushleft}
    \caption{Comparison of inference time \textbf{(left)}, and accuracy loss \textbf{(right)} of \nfgen with \sysname. For both lower is better. 
    We experiment with two approximation error thresholds for \nfgen: i) $\ethreshold =  10^{-1}$ (a crude one we set) and ii) $\ethreshold = 10^{-3}$ (used in~\cite{NFGenFanCCS22}).
     \nfgen ($\ethreshold =  10^{-1}$) achieves a lower inference time but has significant accuracy loss---while \nfgen ($\ethreshold =  10^{-3}$) shows the opposite characteristic when we compare them with \sysname as \#HLs increases. \sysname performs well on both accounts. 
     }
    \label{fig:ablation-study}
    % \fixme{Fix the 2.8$\times$ pos.}
    \vspace{-0.5cm}
  \end{figure}
\changed{\paragraph{Ablation study.}
From the above experiments, it is clear that \sysname is 2$\times$---5$\times$ computationally more efficient than \nfgen while maintaining a negligible accuracy loss.
We perform an ablation study to investigate further what makes \sysname more efficient than \nfgen. For this study, we train $10$ convolutional neural networks on the CIFAR-100 dataset having 1 to 10 hidden layers.
We consider two approximation error thresholds  $\ethreshold$ for  \nfgen: $\ethreshold = 10^{-3}$ as used by Fan et al.~\cite{NFGenFanCCS22}, and $\ethreshold = 10^{-1}$ set by us. \figref{fig:ablation-study} shows the percentage accuracy loss and inference time for 3PC scenario and $\silu$ activation function.     
}

\changed{We observe that, on the one hand, although \nfgen with a crude approximation error threshold $\ethreshold = 10^{-1}$ generates an approximation of $\silu$ that renders inference time closer to \sysname with increasing number of hidden layers, it suffers from significant accuracy loss. On the other hand, \nfgen with $\ethreshold = 10^{-3}$ has negligible accuracy loss similar to \sysname. But it has a high inference time with increasing hidden layers.}

\changed{
We believe this is due to two reasons. First, \sysname automatically finds the optimal $\solution = \langle \npieces, \maxdegree, \ring \rangle$ via binary search over $\ethreshold$ coupled with a simulated annealing-based heuristic. Using a fixed  $\ethreshold$ as \nfgen fails to achieve this. Second,  we incorporate batch normalization into approximation error calculation $\MeanApproxError$~(\eqnref{eqn:max_approximation_error})  to have a better approximation to polynomial pieces near the zero regions. This is different from the approximation error calculated by \nfgen, which uses $\MaxApproxError$. As the number of complex AFs increases from with the number of layers, \sysname's small efficiency for a single complex AF over \nfgen becomes pronounced.}

% Approximation errors in AF amplify intensely as they propagate through subsequent convolution layers. With an increase in the number of layers, this phenomenon repeats, resulting in higher accuracy loss, as mentioned in lines #109-118 and observed in [21, d]. 
% Denoising property of batch normalization could counter some of the errors (e.g., in single-layer DigitRecognition task), but fails when the number of layers is more than three (as shown in Table 3).

\section{Conclusion and Future work} \label{sec:discussion-and-futurework}
\label{sec:discussion}
% \changed{
In this work, we present \sysname that enables fast secure inference for DNN models that use complex activation functions (AFs). We experimentally validate that $\sysname$ have better performance and inference accuracy  on the three most popular complex AFs (i.e., $\silu, \gelu, \mish$) than the state-of-the-art approaches when the number of hidden layers is more than one. Thus, we believe \sysname can accelerate easy adoption of secure inference even when DNN models have a number of hidden layers, trained over complex AFs to generate robust, noise-resistant, better-performing DNN models. 
% (e.g., $\mish$ have higher accuracy in computer vision and object detection tasks than $\relu$~\cite{misra2019mish}, DNN models trained on $\silu$ are more noise-resistant than the ones trained on $\relu$~\cite{hendrycks2016gaussian}, etc.). 
Deploying \sysname is straightforward as it is compatible with standard MPC libraries, and obviates the need to retrain/fine-tune DNN models further after replacing the complex AF with  $\relu$ AF to make it compatible with secure inference protocols specific to $\relu$. Furthermore,
one can easily use our approach to approximate other less widely used complex AFs as well (e.g., $\tanh$, Smish, etc.). 
We point few promising research directions that we hope future work will investigate.
% }
% This part that says we can retrain on relu is implicit. 

\medskip\noindent\textbf{Accelerating secure inference time using GPU.}
In the plaintext setting, impressive ML inference time has been achieved by harnessing 
GPUs, which support highly parallelizable workloads. 
This boost in inference speed can also be extended to secure setting by running MPC operations inside  GPUs. 
Indeed, recent works have shown how to achieve significant speedup in machine learning training and inference by making MPC operations GPU-friendly~\cite{watson2022piranha, tan2021cryptgpu}.
We did not use GPUs while benchmarking secure inference  time. Thus, future work can look at  porting these GPU-friendly techniques to two state-of-the-art secure inference protocols \abythree
and CryptFlow2, we use for benchmarking and improve the inference time reported in~\tabref{tab:performance}.

\paragraph{Dependence on batch normalization.}
As batch normalization (BN) is employed before AFs by the majority of state-of-the-art DNN models used in computer vision, $\sysname$ leverages the phenomenon that BN shifts the input distribution to have zero mean and unit variance. %(similar to standard normal distribution $\normal(0,1)$).  
We believe other  normalization approach, such as layer normalization -- typically used in natural language processing -- % seq2seq model more accurate?
% called layer normalization, 
can also be leveraged to design an approach similar to ours in the future.

% For now, it is clear that if one has a DNN model with a large number of hidden layers, which is already trained using complex AF and batch normalization, % before complex AFs --- 
% \sysname would be useful to do secure inference on that model without sacrificing the accuracy or requiring any retraining. This will significantly simplify deployability. Moreover, compatibility with existing MPC libraries makes it easy for practitioners to easily switch to \sysname if they are already using a secure inference.  

\changed{
\paragraph{Robustness of secure inference and training.}
For safety-critical, and privacy-sensitive applications, understanding the robustness of the model against attacks like training data poisoning, model inversion, adversarial examples, and membership inference attacks is crucial before deploying them in practice.  
% We emphasize that 
% this issue especially because 
To the best of our knowledge, such attacks are not considered by most prior work~\cite{ng2023sok} 
% existing literature 
on secure cryptographic training and inference.
Only recently secure training and inference have 
demonstrated performance levels suitable for practical deployment. Therefore, it is now important to evaluate the robustness of existing secure training or inference protocols against such attacks, before deploying these applications in practice. 
}

\newcommand{\qbb}[1]{\ensuremath \mathbb{Q}_{\mathsf{#1}}}
\newcommand{\dbb}{\ensuremath \mathbb{D}}
\changed{
\paragraph{Membership inference attacks.} \label{app:membership-inference-attacks} 
\sysname uses the training data (or a holdout data) to find an approximated AF that does not degrade accuracy and performance overhead, and one may assume there is a chance that it could affect membership inference attacks for outliers. In fact, any work introducing a cryptography-friendly approximation of AF, that relies on training or holdout data, to balance accuracy and performance, could be susceptible. For example, Delphi~\cite{mishra2020delphi} uses the training data to run a planner that replaces certain $\relu$ AF with cryptography-friendly square function, and  SecureML~\cite{mohassel2017secureml} replaces $\relu$ AF with a new cryptography-friendly approximation. We discuss if using cryptography-friendly approximation of activation functions for secure inference could impact membership inference attacks briefly in~\appref{app:mi}. 
}

\begin{acks}
    We thank the anonymous reviewers for their valuable comments. We also thank Jihye Choi and Maximilian Zinkus for their feedback on the initial draft of this paper.  This work was partially completed when the first and the last authors were at Visa Research.
    This research is supported in part by the University of Wisconsin—Madison Office of the Vice Chancellor for Research and Graduate Education and NSF award \#2339679, and US Department of Commerce award \#70NANB21H043.
\end{acks}

\bibliographystyle{ACM-Reference-Format}
\bibliography{main}

% \begin{appendices}
\appendix
% \section{Performance of Complex Activation Functions over $\relu$} 
% \label{app:performance-of-complex-AF-over-relu}
% \fixme{Discuss about the advantages of using complex AF}

\section{Related Work on Complex Activation Functions} \label{app:related-work}
In this section, we review some work on activation functions (AFs) that fall outside the scope of our work --- but nevertheless be interesting to readers.

\paragraph{Complex AFs used in early NN.}
Early neural network (NN) models 
used binary threshold units~\cite{hopfield1982neural, fitch1944warren} and subsequently $\sigmoid$ and $\tanhh$  
as AFs. The AFs are complex as well. 
However, research in the last decades has exhibited  
$\relu$ AF outperforms such complex used by primary NN models convincingly. Hence, in our work, we focus on more recent complex AFs that have superior performance than $\relu$.  

\paragraph{Linear splines based AFs.}
Methods presented in~\cite{neumayer2022approximation,aziznejad2019deep, bohra2020learning, bohra2021learning}
use a set of linear learnable splines as AFs in 1-Lipschitz constraint neural network. 
These linear splines are MCP-friendly.
But 1-Lipschitz constraint neural network, due to being prone to vanishing gradients problem and having less expressiveness~\cite{li2019preventing}, is not generally used for cloud-based inference services. Thus, they fall outside the scope of our paper.

\paragraph{Polynomial AFs.}
Recently, another trendy line of work attempts to redesign AF with a polynomial function~\cite{diaa2023fast, polykervnets}. During training, these new AF are used instead of the traditional AF. However, in many real-world scenarios, we assume the DNN model has already been trained over traditionally used complex AF, and retraining further is considered expensive and time-consuming. Hence, these works are outside the scope of our work as well.

\changed{\paragraph{Batch normalization and AF approximation.} 
Prior works~\cite{chabanne2017privacy,lopez2019piecewise, wang2022new} that use batch normalization to generate an accurate approximation of activation function mostly focus on $\relu$. Amongst these, only the work from Chabanne et al.~\cite{chabanne2017privacy} falls within the scope of this study. However, their ``least square fit'' approach with a single polynomial makes it unsuitable to render an approximation that has negligible accuracy loss.
% that approach in~\secref{sec:related-work-mpc}. \fixme{Why?}

The other two works, do not fall within the scope, as they require further re-training of the model over the approximated $\relu$ which does not satisfy our design goal \goal{4}.
}

\section{Discussion on Gradient free local optimizations} \label{app:minmax}
% \minote{Needs some care.}
% https://proceedings.mlr.press/v119/liu20j/liu20j.pdf
% We want $\sysname$ to generate MPC-friendly approximation that have both negligible accuracy loss and performance overhead. We treat this problem as a constraint optimization problem (COP), by assigning a constraint how much accuracy loss, say $\negligibleaccuracy$, we can tolerate. More concretely, we can pose this problem as the following equation 
% \changed{
% \bnm 
% \min_{\substack{\solution \in \Theta \\  \fmpc \ \sysname(\solution)}} \Time(\fmpc) \text{ s. t. } \AccuracyLoss(\fmpc) \le \negligibleaccuracy
% % \min_{\substack{c^{(1)}, c^{(2)}, \cdots, c^{(m)}; \\ \mu_1, \mu_2, \cdots, \mu_K}}
% \enm
% }
% \fixme{Need to fix the semantic meaning of this formula}

% Unfortunately, it is not computationally feasible to iterate the search space $\Theta$. 
% Therefore, we look toward constraint local optimizations frameworks that do not require $\Theta$. We also need the local optimization to be gradient-free (or equivalently treat the objective function $\Time$ as a blackbox).

As discussed in~\secref{sec:time}, in this work, we try to solve the constraint optimization problem in~\eqref{eqn:optimization}.

Our initial attempt was to solve~\eqref{eqn:optimization}  
using a hill climbing (HC)~\cite{selman2006hill} type approach. However, it failed as HC tends 
to get stuck while solving optimization problems. 
We, therefore, choose a simulated annealing-based approach since it is scholastic and generally resistant to getting stuck during solution searches. One may explore other techniques from the existing gradient-free local optimization literature (e.g., ant colony optimization, tabu search, genetic algorithm, etc.)~\cite{eren2017introduction}. 
We believe they may work equally well given they are successfully customized for secure inference.

\section{Details of MiniONN} \label{app:MiniONN}
\newcommand{\knots}{S}
\newcommand{\regression}{\funcfont{poly\_regression}}
\newcommand{\fapprox}{\ensuremath \widehat{f}}
\newcommand{\fitGoodness}{\ensuremath \mathsf{fitness}}
\newcommand{\ethresholdmin}{\ethreshold_{\mathsf{min}}}

\begin{figure}[t]
        \fpage{0.45}{
          \underline{\funcfont{MiniONN}~($\npieces, \maxdegree, n, a, b$)}: \\
          \comments{Pick $n$ equally-spaced points in $[a,b]$}\\
          $X \gets \{a+i\cdot(b-a)/n ~|~ i\in \{0,1,\ldots,n\} \}$\\
          % $Y \gets \{\fx(x_i) ~|~ x_i\in X\}$ \comments{$y_i = \fx(x_i)$}\\
          $\knots \gets \{a, b\}$ \\
          \textbf{while} $|\knots| \le m+2$ \textbf{do} \\
          \mytab $x^* \gets \argmin_{\{x\in X\}} \fitGoodness(X, \knots\cup\{x\}, \maxdegree)$\\
          \mytab $S \gets S \cup \{x^*\}$\\
          $\fmpc \gets \emptyset$ \\ 
          \textbf{for} $i \in \{2,\ldots, |S|\}$ \textbf{do} \\
          \mytab $X' \gets \{x \in X\given S[i-1]\le x\le S[i]\}$\\
          \mytab $Y' \gets \{\fx(x)  \given x\in X'\}$ \\  
          \mytab $f_i \gets \mathsf{spline}(X', Y', \maxdegree)$  \\
          \mytab $\fmpc \gets \fmpc \bigcup \{f_i\}$ \\  
          \ret $\fmpc$ \\ 

          $\underline{\fitGoodness(X, S, \maxdegree):}$ \comments{Measures goodness of the fit}\\ 
        \comments{Sort $S$ in ascending order. Also ensure $S[1]=a$ and $S[|S|]=b$}\\
          for $i = 2, \ldots, |S|$ do\\
          \mytab $X' \gets \{x \in X\given S[i-1]\le x\le S[i]\}$\\
          \mytab $Y' \gets \{\fx(x)  \given x\in X'\}$ \\  
          \mytab $f_i \gets \mathsf{polyfit}(X', Y', \maxdegree)$  \\ 
          \mytab $\ethreshold_i \gets \large \sum_{x \in X'} \lVert\fx(x) - f_i(x)\rVert_2$  \\ 
          \ret $\sum_i\ethreshold_i$
        }
        \begin{flushleft}
          \footnotesize
          $^\S$Similar to Liu et al.~\cite{liu2017oblivious}, we use \texttt{scipy.interpolate.UnivariateSpline}, and  \texttt{numpy.polyfit} libraries to implement $\mathsf{Spline}$, and  $\mathsf{polyfit}$ function respectively. 
        \end{flushleft}
  \caption{Recipe used by MiniONN~\cite{liu2017oblivious} to approximate activation functions $\fx$ between $[a,b]$ by a set of $\npieces$ piece-wise continuous splines $\fmpc = \bigcup_{i=1}^{\npieces} \{\poly_{i}\}$ such that  $\forall i \ \deg(\poly_i) \le \maxdegree$. 
  % \fixme{recheck this}
  }
  \label{fig:MiniONN}
\end{figure}

Liu et al.~\cite{liu2017oblivious} proposed MiniONN, a scheme to 
to generate MPC-friendly piece-wise polynomials approximations for 
$\sigmoid$, and $\mathsf{tanh}$ function.
We  implement this recipe ourselves to generate MPC-friendly approximations for complex activation functions (AFs) (e.g., $\silu$, $\gelu$, $\mish$) between range $a = -5$, and $b = 5$ as shown in~\figref{fig:MiniONN}

Briefly, this approach takes $n$ equally spaced points between $[a,b]$ to approximate the given  AF $\fx$. Let's denote the set of these points as
$X \gets \{x_1, x_2, \ldots, x_n\}$ where $x_1 = a$, and $x_n = b$, and  $Y \gets \{ y_1, y_2, \ldots, y_n\}$ be the set of values used to approximate $\fx$ where $y_i = \fx(x_i)$. Then MiniONN attempts to find a set of $m$ switchover points $S \gets \{s_1, s_2, \ldots s_m\}$ between $[a,b]$. The points from $S$ are used as knots to approximate 
$\fx$ using $m+1$ MPC-friendly polynomials (same as~\eqnref{eqn:approx}); 
where $\poly_i$ is a spline approximating the region between 
$\{s_i, s_{i+1}\}$ for $i \in \{1,2, \ldots m-1\}$. Note $\poly_0 = 0$, and $\poly_{m+1}(x) = x$.
MiniONN finds these $m$ switching points by iterating over each $x \in X$ and selecting a new point for iteration that maximizes the overall goodness of the fit. 

For our case, we consider squared mean error as way to measure the goodness of the fit --- as shown in the $\fitGoodness$ procedure~(\figref{fig:MiniONN}). We explored a few parameters for $\langle n, m, k \rangle$ and settle for 
$n = 1,000$, $m = 20$,  and $\maxdegree = 3$ since it 
yields the best accuracy as reported in~\tabref{tab:classification-accuracy}.

\section{DNN Models Considered for Different Tasks}
\label{app:dnn}
\begin{figure}[t]
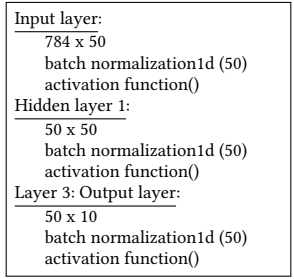

    \centering
    \fpage{0.20}{
        \underline{Input layer}: \\ 
        \mytab 784 x 50 \\ 
        \mytab batch normalization1d (50) \\ 
        \mytab activation function() \\ 
        \underline{Hidden layer 1}: \\ 
        \mytab 50 x 50 \\ 
        \mytab batch normalization1d (50) \\ 
        \mytab activation function() \\
        \underline{Layer 3: Output layer}: \\ 
        \mytab 50 x 10 \\ 
        \mytab batch normalization1d (50) \\ 
        \mytab activation function()  
    }
\caption{Fully connected neural network with one hidden layer we use for \taskA{} task.}
\label{fig:fcn}
\end{figure}

\begin{figure}[t]
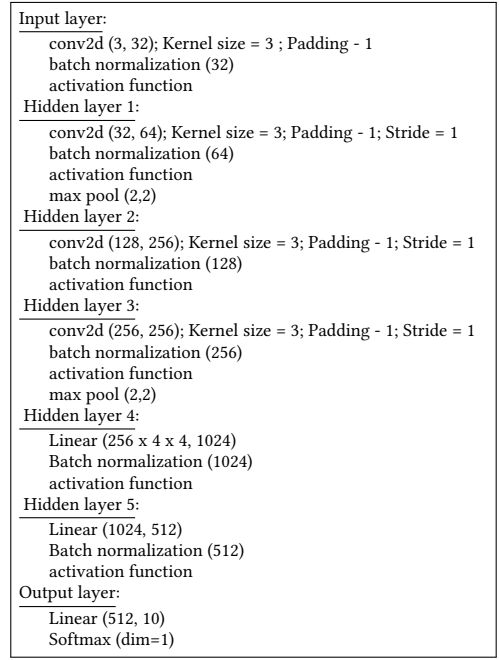

    \centering
        \fpage{0.35}{
            \underline{Input layer}: \\ 
            \mytab conv2d (3, 32); Kernel size = 3 ; Padding - 1\\
            \mytab batch normalization (32) \\ 
            \mytab activation function \\ 
            \underline{ Hidden layer 1}: \\ 
            \mytab conv2d (32, 64); Kernel size = 3; Padding - 1; Stride = 1\\
            \mytab batch normalization (64) \\ 
            \mytab activation function \\
            \mytab max pool (2,2) \\  
            \underline{ Hidden layer 2}: \\ 
            \mytab conv2d (128, 256); Kernel size = 3; Padding - 1; Stride = 1\\
            \mytab batch normalization (128) \\ 
            \mytab activation function \\
            \underline{ Hidden layer 3}: \\ 
            \mytab conv2d (256, 256); Kernel size = 3; Padding - 1; Stride = 1\\
            \mytab batch normalization (256) \\ 
            \mytab activation function \\
            \mytab max pool (2,2) \\
            \underline{ Hidden layer 4}: \\ 
            \mytab Linear (256 x 4 x 4, 1024) \\
            \mytab Batch normalization (1024) \\
            \mytab activation function \\
            \underline{ Hidden layer 5}: \\
            \mytab Linear (1024, 512) \\ 
            \mytab Batch normalization (512) \\
            \mytab activation function \\
            \underline{Output layer}: \\
            \mytab Linear (512, 10) \\ 
            \mytab Softmax (dim=1) 
        }
  \caption{Convolutional neural network (CNN) model with 5 hidden layers we use for \taskB{} task.}
  \label{fig:convnet}
\end{figure}

% \fixme{There are tools to draw DNN models. Use those}
For experimental evaluation, in this work, we consider four  
DNN models for four image classification tasks.
For \taskA task, we consider a fully connected neural network (FCN) with one hidden layer as shown in~\figref{fig:fcn}.
For the second task \taskB, we consider a convolutional neural network (CNN) with five hidden layers (\figref{fig:convnet}).
Since the datasets we consider for the last two tasks are relatively more complex, we select complex DNN models for them. 

\begin{figure}[t]
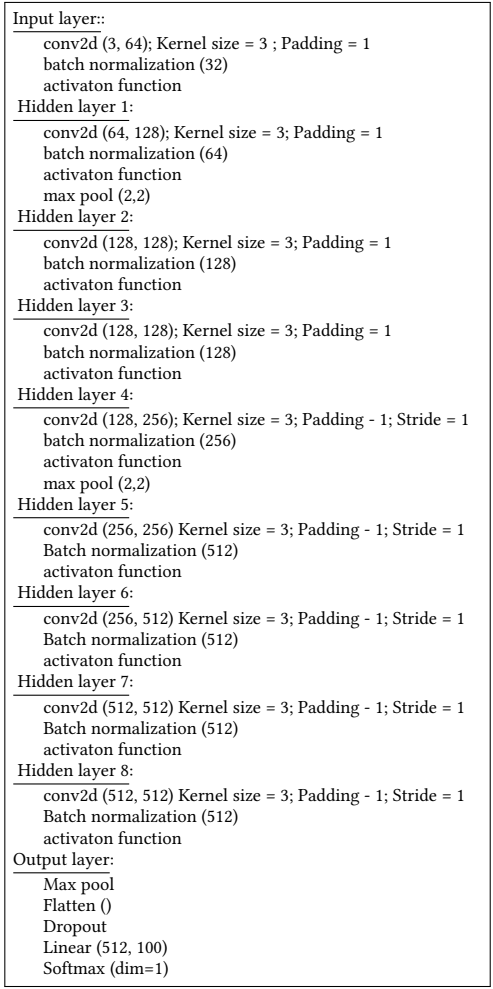

    \fpage{0.35}{
        \underline{Input layer:}: \\ 
        \mytab conv2d (3, 64); Kernel size = 3 ; Padding = 1\\
        \mytab batch normalization (32) \\ 
        \mytab activaton function \\ 
        \underline{ Hidden layer 1}: \\ 
        \mytab conv2d (64, 128); Kernel size = 3; Padding = 1\\
        \mytab batch normalization (64) \\ 
        \mytab activaton function \\
        \mytab max pool (2,2) \\  
        \underline{ Hidden layer 2}: \\ 
        \mytab conv2d (128, 128); Kernel size = 3; Padding = 1\\
        \mytab batch normalization (128) \\ 
        \mytab activaton function \\
        \underline{ Hidden layer 3}: \\ 
        \mytab conv2d (128, 128); Kernel size = 3; Padding = 1\\
        \mytab batch normalization (128) \\ 
        \mytab activaton function \\
        \underline{ Hidden layer 4}: \\ 
        \mytab conv2d (128, 256); Kernel size = 3; Padding - 1; Stride = 1\\
        \mytab batch normalization (256) \\ 
        \mytab activaton function \\
        \mytab max pool (2,2) \\
        \underline{ Hidden layer 5}: \\ 
        \mytab conv2d (256, 256)  Kernel size = 3; Padding - 1; Stride = 1\\
        \mytab Batch normalization (512) \\
        \mytab activaton function \\
        % \underline{Layer 7: convolutional layer}: \\ 
        % \mytab conv2d(256, 256)  Kernel size = 3; Padding - 1; Stride = 1\\
        % \mytab Batch normalization (512) \\
        % \mytab activaton function \\
        \underline{ Hidden layer 6}: \\ 
        \mytab conv2d (256, 512)  Kernel size = 3; Padding - 1; Stride = 1\\
        \mytab Batch normalization (512) \\
        \mytab activaton function \\
        \underline{ Hidden layer 7}: \\ 
        \mytab conv2d (512, 512)  Kernel size = 3; Padding - 1; Stride = 1\\
        \mytab Batch normalization (512) \\
        \mytab activaton function \\
        \underline{ Hidden layer 8}: \\ 
        \mytab conv2d (512, 512)  Kernel size = 3; Padding - 1; Stride = 1\\
        \mytab Batch normalization (512) \\
        \mytab activaton function \\
        \underline{Output layer}: \\
        \mytab Max pool \\ 
        \mytab Flatten () \\ 
        \mytab Dropout  \\ 
        \mytab Linear (512, 100) \\ 
        \mytab Softmax (dim=1) 
    }
\caption{ResNet9~\cite{he2016deep} DNN architecture with eight hidden layers we use for \taskC{} task.}
% \fixme{Draw this as figures.}
% https://1.bp.blogspot.com/-DjZT_TLYZok/XO3BYqpxCJI/AAAAAAAAEKM/BvV53klXaTUuQHCkOXZZGywRMdU9v9T_wCLcBGAs/s640/image2.png
% https://stackoverflow.com/questions/73308149/meaning-of-output-shapes-of-resnet9-model-layers
\label{fig:resnet9}
\end{figure}

\begin{figure}[t]
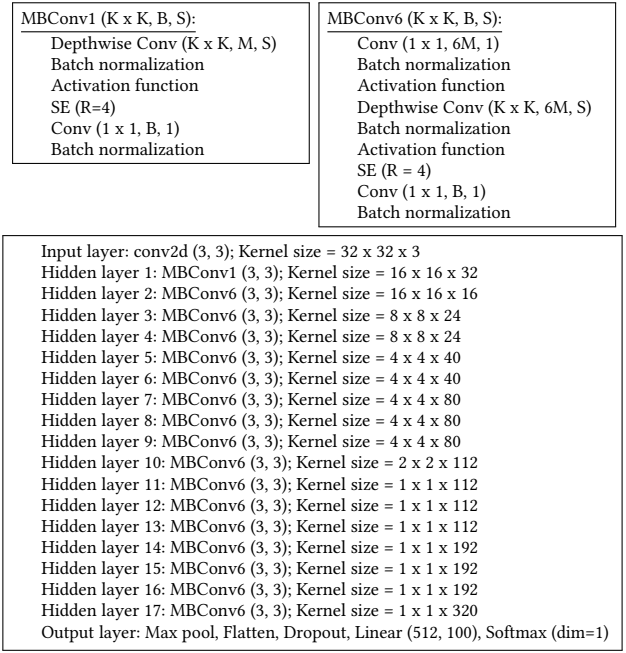

    \hpagess{0.22}{0.22}{
    \fpage{0.95}{
    \underline{MBConv1 (K x K, B, S):} \\ 
        \mytab Depthwise Conv (K x K, M, S) \\ 
        \mytab  Batch normalization \\ 
        \mytab  Activation function \\ 
        \mytab  SE (R=4) \\ 
        \mytab  Conv (1 x 1, B, 1) \\ 
        \mytab Batch normalization 
    }}
    {
    \fpage{0.95}{
    \underline{MBConv6 (K x K, B, S):} \\
        \mytab Conv  (1 x 1, 6M, 1) \\ 
        \mytab  Batch normalization \\ 
        \mytab  Activation function \\ 
        \mytab  Depthwise Conv (K x K, 6M, S) \\ 
        \mytab  Batch normalization \\ 
        \mytab Activation function \\
        \mytab SE (R = 4) \\ 
        \mytab Conv (1 x 1, B, 1) \\ 
        \mytab Batch normalization
    }}\\ \smallskip
    \fpage{0.45}{
    % \underline{EfficientNetB0 Model:} \\
        \mytab Input layer:  conv2d (3, 3); Kernel size = 32 x 32 x 3 \\
        \mytab Hidden layer 1: MBConv1 (3, 3); Kernel size = 16 x 16 x 32 \\
        \mytab Hidden layer 2: MBConv6 (3, 3); Kernel size = 16 x 16 x 16 \\
        \mytab Hidden layer 3: MBConv6 (3, 3); Kernel size = 8 x 8 x 24 \\
        \mytab  Hidden layer 4: MBConv6 (3, 3); Kernel size = 8 x 8 x 24 \\
        \mytab  Hidden layer 5: MBConv6 (3, 3); Kernel size = 4 x 4 x 40 \\
        \mytab  Hidden layer 6: MBConv6 (3, 3); Kernel size = 4 x 4 x 40 \\
        \mytab  Hidden layer 7: MBConv6 (3, 3); Kernel size = 4 x 4 x 80 \\
        \mytab  Hidden layer 8: MBConv6 (3, 3); Kernel size = 4 x 4 x 80 \\
        \mytab  Hidden layer 9: MBConv6 (3, 3); Kernel size = 4 x 4 x 80 \\
        \mytab  Hidden layer 10: MBConv6 (3, 3); Kernel size = 2 x 2 x 112 \\
        \mytab  Hidden layer 11: MBConv6 (3, 3); Kernel size = 1 x 1 x 112 \\
        \mytab  Hidden layer 12: MBConv6 (3, 3); Kernel size = 1 x 1 x 112 \\
        \mytab  Hidden layer 13: MBConv6 (3, 3); Kernel size = 1 x 1 x 112 \\
        \mytab  Hidden layer 14: MBConv6 (3, 3); Kernel size = 1 x 1 x 192 \\
        \mytab  Hidden layer 15: MBConv6 (3, 3); Kernel size = 1 x 1 x 192 \\
        \mytab  Hidden layer 16: MBConv6 (3, 3); Kernel size = 1 x 1 x 192 \\
        \mytab  Hidden layer 17: MBConv6 (3, 3); Kernel size = 1 x 1 x 320 \\ 
        \mytab Output layer: Max pool, Flatten, Dropout, Linear (512, 100), Softmax (dim=1)
    } 

\caption{EfficientNetB0~\cite{tan2019efficientnet} DNN architecture with 17 
% eight hidden 
layers we use for \taskD task.}
\label{fig:efficientNetB0Model}
\end{figure}

In particular, for \taskC, we consider a residual neural network (ResNet9) having eight hidden layers (\figref{fig:resnet9}). For \taskD task we choose the EfficientNetB0 model having 17 hidden layers (\figref{fig:efficientNetB0Model}). 
We refer the interested reader to~\cite{he2016deep, tan2019efficientnet} for more details. Reader should observe that inputs to the activation function are batch normalized for each of the four models -- a standard practice in DNN models.

% \begin{figure*}
%     \centering
%     \includegraphics{figures/models/efficientNetB0.jpg}
%     \caption{EfficientNetB0 model with 17 hidden layers. Image is taken from~\cite{gang2021character}\fixme{Redraw this on your own}.}
%     \label{fig:efficientNetB0Model}
% \end{figure*}

\section{2PC Results} \label{app:crytpflow2}
\begin{table}[t]
    \centering 
    \tabfontsize
    \begin{tabular}{@{}lrlrrr@{}}
        \toprule
        Task Name  & \# HLs$^\dag$ & $\fx$ & NFGen & Ours & \textbf{Speedup} \\ \midrule 
         \multirow{3}{*}{\taskA} & \multirow{3}{*}{1} & $\mish$ &  $184$ & $170$ & $1.08\times$\\
         & & $\silu$ &  $290$ & $305$ & $0.95\times$ \\ 
         & & $\gelu$ &    $135$ & $133$ & $1.01\times$\\ \midrule 
         \multirow{3}{*}{\taskB} & \multirow{3}{*}{5} & $\silu$ &  $723$ & $205$ & \textbf{$3.52\times$} \\ 
         & & $\gelu$ & $745$ & $165$ & $4.50\times$ \\
         & & $\mish$ & $825$ & $229$ & $3.60\times$ \\ \midrule 
         \multirow{3}{*}{\taskC} & \multirow{3}{*}{8} & $\silu$ & $512$ & $122$ & \textbf{$4.20\times$}\\ 
         & & $\gelu$ & $502$ & $108$ & $4.64\times$\\ 
         & & $\mish$ & $537$ & $147$ & $3.63\times$\\  \midrule
        \multirow{3}{*}{\taskD} & \multirow{3}{*}{17} & $\silu$ & $827$ & $189$ & $4.37\times$\\ 
         & & $\gelu$ & $876$ & $192$ & $4.54\times$ \\ 
         & & $\mish$ & $893$ & $203$ & $4.38\times$\\ 
        \bottomrule
\end{tabular}

    \begin{flushleft}
    $^\dag$ \# HLs =  Number of hidden layers. We experiment with the oblivious transfer (OT) based construction of CryptFlow2. 
    \end{flushleft}
    \caption{
    Comparison of inference time (ms) of three activation functions ($\fx$) over four different classification tasks for $N=2$ servers using CryptFlow2 MPC library. Since the  DNN model used in $\taskA{}$ task has only one hidden layer (\# HLs=1), the performance of $\nfgen$  task is similar to \sysname. However, as DNN models become more complex and deep, having high hidden layers; for the other three tasks \sysname outperforms $\nfgen$ --- exhibiting a speedup $2\times$--$5\times$ when compared to $\nfgen$.
    }
        \label{tab:performance-appendix}

\end{table}
We use the oblivious transfer-based construction for CryptFlow2. We observe a similar performance gain as the 3PC scenario --- 2$\times$-5$\times$ speedup of \sysname compared to \nfgen --- as shown in~\tabref{tab:performance-appendix}.

\begin{table}[t]
    \centering
    \tabfontsize
    \begin{tabular}{@{}lllrr@{}}
        \toprule 
        Task Name & $\fx$ & $\npieces$ & $\maxdegree$ & $\ring$ \\  
        \midrule
        \multirow{3}{*}{\taskA} & $\silu$ & 78 & 3 & $\langle 64,32 \rangle$\\
         & $\gelu$ & 82 & 5 & $\langle 64,32 \rangle$\\
         & $\mish$ & 85 & 3 & $\langle 64,32 \rangle$\\ 
         \midrule
        \multirow{3}{*}{\taskB} & $\silu$ & 77 & 3 & $\langle 84, 42 \rangle$\\
         & $\gelu$ & 34 & 3 & $\langle 84, 63 \rangle$\\
         & $\mish$ & 93 & 4 & $\langle 84, 42 \rangle$\\
         \midrule
        \multirow{3}{*}{\taskC} & $\silu$ & 91 & 5 & $\langle 84, 42 \rangle$\\
         & $\gelu$ & 101 & 5 & $\langle 84, 42\rangle$ \\
         & $\mish$ & 96 & 5 & $\langle 84, 42 \rangle$ \\
         \midrule
        \multirow{3}{*}{\taskD} & $\silu$ & 81 & 5 & $\langle 64, 32 \rangle$ \\
         & $\gelu$ & 90 & 5 & $\langle 64, 32 \rangle$ \\ 
         & $\mish$ & 93 & 5 & $\langle 64, 32\rangle$\\ 
        \bottomrule 
    \end{tabular}
    \caption{$\npieces, \maxdegree, \ring$  for four tasks using $\ApproximationAlgo$.
    We consider accuracy loss $\negligibleaccuracy < 10^{-2}$ as negligible. $\ring$ is presented as $\langle \resolution,\decimal \rangle$. }
    \label{tbl:configurations}
    \vspace{-1em}
\end{table}

\section{Membership Inference attacks} \label{app:mi}
Membership inference (MI) is a popular way to 
examine the privacy of the training data and
predicts if  a given input $(x,y)$ was used to train a model $f_{\theta}$. We focus on the  state-of-the-art MI attack by Carlini et al.~\cite{carlini2022membership}. 

To find if a specific input value $(x,y)$ was used to train a model, the attacker first creates $N$ samples of training data from the data distribution $\dbb$, such that half of the datasets contain $(x,y)$ in them, and other half do not. Then the attacker trains two sets of shadow models $M_1$ (where the datasets contain $(x,y)$) and $M_2$ (where the datasets do not contain $(x,y)$). 
% on this training data shadow models $M_1$ and $M_2$ on random training data : % Here $N$ is the number of shadow models attackers want to train.
% each model in $M_1$ is trained over different random datasets $D$, such that each $D\cup\{(x,y)\}$ where $D\getsr \dbb$, and  $M_2$ is trained over $\tau(D \setminus \{x, y\} | D \getsr \dbb)$ where $\dbb$ is the distribution of the data.
$\qbb{in}$ denotes the distribution  of losses of $(x,y)$ from $M_1$ and $\qbb{out}$ is the distribution  of the cross losses from $M_2$.
Finally, the attacker calculates the cross entropy loss of $(x,y)$ on the target model $\ell(f_\theta(x),y)$ and measure the likelihood of this loss under the distributions $\qbb{in}(x,y)$ and $\qbb{out}(x,y)$ and return whichever is more likely.

% $M_1$ is trained over  $\tau(D \cup \{x, y\} | D \getsr \dbb)$, and  $M_2$ is trained over $\tau(D \setminus \{x, y\} | D \getsr \dbb)$ where $\dbb$ is the distribution of the data. $\qbb{in}$ denotes the distribution  of losses of  $(x,y)$ from $M_1$ and $\qbb{out}$ is the distribution  of the cross losses from $M_2$.
% Finally, we calculate the cross entropy loss of $(x,y)$ on the target model $\ell(f_\theta(x),y)$ and measure the likelihood of this loss under the distributions $\qbb{in}(x,y)$ and $\qbb{out}(x,y)$ and return whichever is more likely.  

Thus, any approach~\cite{mishra2020delphi,mohassel2017secureml} (including ours) leveraging training data (or a holdout data) to find an approximate AF % that maintains negligible accuracy loss and performance overhead  
could affect $\ell(f_\theta(x),y)$, $\qbb{in}$, and $\qbb{out}$, and thus potentially has an impact on MI attack.
However,  since all these prior works have negligible accuracy loss over the testing data which is sampled from the distribution $\dbb$ (i.e., in-distribution-data), we conjecture that
there is a chance that $\ell(f_\theta(x),y)$, and $\qbb{in}$ or $\qbb{out}$ remain mostly similar when the given input $\{x, y\}$ is sampled from the in-distribution data, and thus may not affect MI attack.

When the given input  $\{x, y\}$ is sampled from outside the in-distribution data (i.e., outliers), it is  difficult to assess how $\ell(f_\theta(x),y)$,  $\qbb{in}(x,y)$ or $\qbb{out}(x,y)$ would get affected.
Interestingly, Carlini et al.~\cite{carlini2022membership} briefly discuss that outliers are inherently more vulnerable to their MI attack. This is because, as they experimentally show, the gap between $\qbb{in}$ and $\qbb{out}$ widens when outliers are inserted into the training dataset of $f_{\theta}$~\cite[Figure 11]{carlini2022membership}, and this enables the attacker to make predictions about $\{x,y\}$ with higher accuracy.

In summary, how  $\ell(f_\theta(x),y)$, and $\qbb{in}$, or $\qbb{out}$ are affected for both in-distribution, and outlier data due to the introduction of secure inference or training protocols, and whether such secure protocols can be used to defend against MI attacks is an interesting open question.

% \newpage 
% \thispagestyle{empty}
% \newpage
% \input{sections/summary-of-changes.tex}

\end{document}